\documentclass[apj]{emulateapj}
\usepackage{longtable}

\newcommand{\cena}{NGC~5128}
\newcommand{\etal}{et~al.\ }

\newcommand{\perpix}{{\rm pixel^{-1}}}
\newcommand{\lya}{Lyman~$\alpha \ $}
\newcommand{\Ha}{H$\alpha$}      
\newcommand{\OIII}{\ion{O}{3}}   
\newcommand{\kms}{km~s$^{-1}$}

\newcommand{\uv}{$U$--$V$}

\journalinfo{Astrophysical Journal, v602, 2004 February 20}
\submitted{Received 2003 August 14; accepted 2003 November 4}

\shorttitle{Planetary Nebulae in \cena}
\shortauthors{Peng, Ford, \& Freeman}

\begin{document}


\title{The Planetary Nebula System and Dynamics in the Outer Halo of \cena}


\author{Eric W.\ Peng\altaffilmark{1,2}, Holland C.\ Ford\altaffilmark{1,3}}
\affil{Department of Physics and Astronomy, Johns Hopkins
        University, Baltimore, MD, 21218, USA}
\email{ericpeng@pha.jhu.edu, ford@pha.jhu.edu}

\and

\author{Kenneth C.\ Freeman}
\affil{RSAA, Australian National University, Canberra, ACT, Australia}
\email{kcf@mso.anu.edu.au}


\altaffiltext{1}{Visiting Astronomer, Cerro Tololo Inter-American Observatory,
which is operated by the Association of Universities for Research in
Astronomy, Inc.\ (AURA) under cooperative agreement with the National Science
Foundation.}
\altaffiltext{2}{Current address: 136 Frelinghuysen Road, Physics and
Astronomy, Rutgers University, Piscataway, NJ 08854, USA;
ericpeng@physics.rutgers.edu}
\altaffiltext{3}{Space Telescope Science Institute, 3700 San Martin Drive,
        Baltimore, MD 21218, USA}


\begin{abstract}
The halos of elliptical galaxies are faint and difficult to explore, but
they contain vital clues to both structure and formation.  
We present the results of an imaging and spectroscopic survey for planetary
nebulae (PNe) in the nearby elliptical \cena.  We extend the work
of Hui \etal (1995) well into the halo of the galaxy---out to distances
of 100 and 50~kpc along the major and minor axes.  
We now know of 1141 PNe in \cena, 780 of which are
confirmed.  Of these 780 PNe, 349 are new from this survey, and 148 are
at radii beyond 20~kpc.  PNe exist at distances up to 80~kpc ($\sim15r_e$), 
showing that the stellar halo extends to the limit of our data.
This study represents by far the largest kinematic study of an
elliptical galaxy to date, both in the number of velocity tracers
and in radial extent.
We confirm the large rotation of the PNe along the major axis, and show that it
extends in a disk-like feature into the halo.  The rotation curve of the
stars flattens at $\sim100$~\kms\ with $V/\sigma$ between 1 and 1.5, and
with the velocity dispersion of the PNe falling gradually at larger radii.
The two-dimensional velocity field exhibits a zero-velocity
contour with a pronounced twist, showing that the galaxy potential is
likely triaxial in shape, tending toward prolate.  The total dynamical
mass of the
galaxy within 80~kpc is $\sim5\times10^{11} M_{\sun}$, with $M/L_B\sim13$.
This mass-to-light ratio is much lower than what is typically expected for
elliptical galaxies.
\end{abstract}




\section{Introduction}

\begin{deluxetable*}{llll}
\tablewidth{0pt}
\tablecaption{Observing Runs: Imaging \label{table:obsrun_im}}
\tablehead{
\colhead{Data} & \colhead{Acquired AT/With} &
\colhead{Date} 
}
\startdata
$\!$ [\ion{O}{3}] on-/off-band, inner halo & CTIO, Prime Focus CCD & 
1993 Jun 17--19 \\
$\!$ [\ion{O}{3}] on-/off-band, outer halo & CTIO, Prime Focus CCD & 
1995 Mar 24--28 \\
$\!$ [\ion{O}{3}] and {\it V} imaging & CTIO, Mosaic II CCD camera &
2000 Jun 4 \\ 
\enddata
\end{deluxetable*}

Extragalactic planetary nebulae (PNe) are powerful tools for studying
the dynamical states of early-type galaxies.  Using their bright
[\ion{O}{3}] emission lines, we can identify PNe and typically
measure their radial velocities to an accuracy of $\sim15\,{\rm km
\,s^{-1}}$.  Traditionally, the kinematics of old stellar populations
are derived from absorption line profiles in integrated spectroscopy
(e.g.\ Bender, Saglia, \& Gerhard 1994; Kronawitter \etal 2000).
This method, however, is limited to the relatively high surface
brightness central regions of galaxies ($r < 2r_e$), where the mass
density is dominated by baryons.  Unlike in spirals where
the rotation curves can be measured out to a large radii using
\ion{H}{1} observations, very few ellipticals have large scale
\ion{H}{1} disks or polar rings that can be used to probe the
potential in the halo.  So, while current theories of structure
formation predict that elliptical galaxies should reside in
massive dark matter halos, the dynamical evidence for these halos is
not nearly as complete or as conclusive as it is for spiral galaxies
(e.g.\ Gerhard \etal 2001, Romanowsky \etal 2003).

Using PNe, we can now obtain kinematical information in the halos of 
early-type galaxies at radii up to 80~kpc, allowing us to study the 
two-dimensional stellar kinematics and the mass distribution in these galaxies.
Because PNe are evolved asymptotic giant branch (AGB) stars, their
kinematics must trace that of the underlying stellar population,
making them ideal test particles for dynamical studies.
PNe have been used successfully as velocity test particles in many
elliptical galaxies.  As early as a decade ago, Ciardullo \etal (1993)
measured velocities for 29 PNe within $r\sim10$~kpc for NGC~3379.  
Other examples include Arnaboldi \etal (1998), who discovered 43 PNe within
$r\sim16$~kpc for NGC~1316 (Fornax A), and M{\'e}ndez \etal (2001), who
measured velocities for 535 PNe within $r\sim14$~kpc for NGC~4697. 
More recently, Romanowsky \etal (2003) presented results from a PN
spectrograph study in which $\sim100$~PNe were observed in each of three
galaxies out to radii of 4--6~$r_e$.

The work most relevant to this paper is the study of Hui \etal
(1995, hereafter H95), who first investigated the PN system of \cena.
At a distance of 3.5 Mpc (Hui \etal 1993a, hereafter H93a), NGC 5128 is the
closest large elliptical galaxy to the Milky Way and is a prime candidate
for PNe studies.  At the distance of \cena, $1\arcmin=1.02 \ {\rm kpc}$.
Hui \etal 1993b (H93b) identified 785 PNe in \cena, and H95 measured
velocities for 431 PNe.  This previous study probed the halo potential
out to projected distances
of $\sim20\,{\rm kpc}$ along the major axis and $\sim10\,{\rm kpc}$
along the minor axis, and produced strong evidence for: 1)~A dark matter
halo, 2)~A significant quantity of angular momentum in the halo, and
3)~Intrinsic triaxiality in the true shape of the galaxy.

In this study, we present the results of an
extended survey in which we have identified and measured velocities for
PNe in \cena\ out to projected distances of 80 and 40 kpc along 
the major and minor axes, respectively.  
We identified 356 new PN candidates, many of which are beyond 20~kpc,
bringing the total number of PNe in \cena\ to 1141.
In total, we were able to measure 349 new PN radial velocities, which
increases the kinematic sample to 780 PNe.  
Similar studies in other galaxies have typically used 2--10 times fewer
test particles (whether PNe or globular clusters), and have mostly been
conducted in regions where $r\lesssim 5 r_e$.  The data set we present in this
paper represents by far the largest kinematic survey of an elliptical galaxy
to date, both in the number of velocity tracers, and the radial extent
in the galaxy halo ($r\sim15 r_e$).

\section{The Survey \label{sec:survey}}

We can identify extragalactic PNe candidates by their bright [\ion{O}{3}] 
emission lines.  PNe, which have strong emission lines and
very little continuum emission, 
are detected in images taken through an ``on-band'' [\ion{O}{3}] filter, 
but are not detected in images taken through an ``off-band''
continuum filter.  
Optimally, the on-band filter is designed such that its central
wavelength matches the wavelength of
$\lambda5007$ redshifted to the systemic velocity of the galaxy,
and its width is large enough to encompass the velocity dispersion
of the observed galaxy's stars.  For \cena, its systemic velocity of
541~\kms\ (H95) translates into an on-band filter centered at 5016\AA.  A
typical galactic velocity dispersion of 200~\kms\ requires a filter width
of greater than 20\AA\ in order to include all PNe with velocities of
less than 3-$\sigma$ from the mean.
Ideally, the off-band filter is an intermediate-band filter that is not
far from the [\ion{O}{3}] line, but does not include it.  In some cases,
a Johnson {\it V}-band filter is used as the off-band.  
The {\it V} filter can be used as an off-band despite the fact
that it contains the [\ion{O}{3}] line because of its large bandpass
($\sim900$\AA).

After identifying PNe candidates in the imaging, we obtain spectra of
them with multi-fiber spectrographs.  The detection 
of the [\ion{O}{3}]$\lambda5007$ emission line allows us to
measure their radial velocities.  In bright PNe, we
can also see the [\ion{O}{3}]$\lambda4959$ emission line.
For a small subset of our sample (22 PNe), we measured the velocity 
using the \Ha\ emission line.

\subsection{Imaging}
\subsubsection{Observations}

\begin{deluxetable}{cccc}[b]
\tablewidth{0pt}
\tabletypesize{\scriptsize}
\tablecaption{Imaged Fields \label{table:fields}}
\tablehead{
\colhead{Field} & \colhead{RA(2000)} & \colhead{Dec(2000)} &
\colhead{Field-of-View (\arcmin)}
}
\startdata
f10 & 13:23:21.26 & $-$43:15:20.09 & $15.1\times14.9$ \\
f12 & 13:27:14.46 & $-$43:01:18.30 & $15.1\times14.9$ \\
f14 & 13:24:36.65 & $-$43:01:01.59 & $15.1\times14.9$ \\
f16 & 13:28:30.72 & $-$42:46:55.61 & $15.1\times14.9$ \\
f17 & 13:26:46.06 & $-$42:47:00.56 & $15.1\times14.9$ \\
f18 & 13:25:56.82 & $-$42:46:56.58 & $15.1\times14.9$ \\
f21 & 13:28:03.09 & $-$42:33:08.67 & $15.1\times14.9$ \\
f8 & 13:25:27.89 & $-$43:15:41.55 & $15.1\times14.9$ \\
f22 & 13:26:44.92 & $-$42:32:47.81 & $15.1\times14.9$ \\
\tableline
f15 & 13:22:54.30 & $-$43:00:41.51 & $14.7\times14.7$ \\
f19 & 13:24:07.66 & $-$42:46:45.31 & $14.7\times14.7$ \\
f20 & 13:22:55.67 & $-$42:46:51.95 & $14.7\times14.7$ \\
f21 & 13:27:58.66 & $-$42:32:49.00 & $14.7\times14.7$ \\
f23 & 13:25:27.95 & $-$42:33:00.94 & $14.7\times14.7$ \\
f24 & 13:24:11.27 & $-$42:33:00.52 & $14.7\times14.7$ \\
f30 & 13:24:09.04 & $-$43:42:59.67 & $14.7\times14.7$ \\
f31 & 13:22:48.88 & $-$43:42:44.84 & $14.7\times14.7$ \\
f40 & 13:24:12.59 & $-$42:18:58.32 & $14.7\times14.7$ \\
f41 & 13:25:28.32 & $-$42:18:53.75 & $14.7\times14.7$ \\
f42 & 13:26:43.04 & $-$42:18:48.18 & $14.7\times14.7$ \\
f43 & 13:27:59.18 & $-$42:18:56.95 & $14.7\times14.7$ \\
f5 & 13:22:54.56 & $-$43:28:48.03 & $14.7\times14.7$ \\
f7 & 13:26:44.96 & $-$43:14:48.87 & $14.7\times14.7$ \\
f4 & 13:24:08.97 & $-$43:28:55.40 & $14.7\times14.7$ \\
\tableline
mosNE & 13:28:13.59 & $-$42:04:24.82 & $38.1\times43.8$ \\
mosSW & 13:21:54.56 & $-$44:06:50.84 & $37.1\times37.6$ \\
\enddata
\end{deluxetable}

\begin{figure*}
\centerline{\includegraphics[width=5.9in]{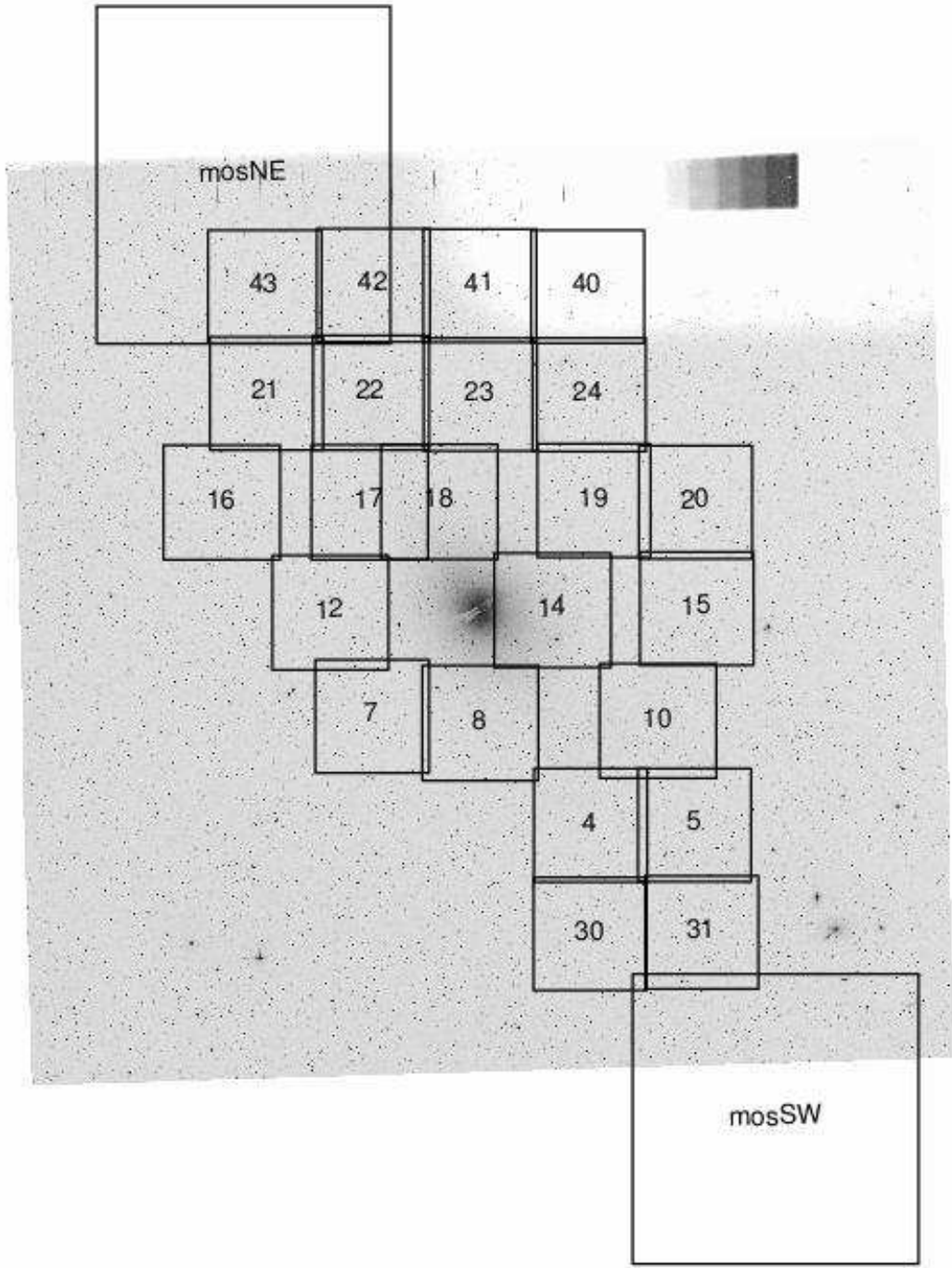}}
\caption[Fields imaged for NGC 5128 PNe Survey]
{\small
Fields imaged for NGC 5128 PNe Survey.  The fields
imaged for PNe detection are plotted over a $2\degr\times2\degr$ 
Digitized Sky Survey image of NGC 5128.  The image is oriented with
North up, East to the left.  The calibration wedge and vignetting
at the top are due to this region of sky being near the edge of the
plate.  At the distance of \cena, $1\arcmin=1.02\ {\rm kpc}$.  The PFCCD
fields are $\sim15\arcmin$ square.  The larger Mosaic II fields are
$\sim38\arcmin$ square. See Table~\ref{table:fields} for central
coordinates and exact sizes for each field.  \label{figure:pnfields}}
\end{figure*}

We imaged the halo of \cena\ over the course of three observing runs
using the Blanco 4-meter telescope at Cerro Tololo Inter-American
Observatory (CTIO).  These imaging runs are listed in
Table~\ref{table:obsrun_im}.  
All imaged fields are listed in Table~\ref{table:fields}, and shown on a
$2\degr\times2\degr$ image of \cena\ from the Digitized Sky Survey in
Figure~\ref{figure:pnfields}.

During the first run, on 1993 Jun 17--19, we
used the Prime Focus CCD Imager (PFCCD) to image twelve halo fields.  
In 1993, the PFCCD $2048\times2048$ pixel
detector covered a $15\farcm1\times14\farcm9$ field on the sky at a
pixel scale of $0\farcs445 \ \perpix$.  Typical seeing was in the range 
$1\farcs2$--$1\farcs7$.  There was thin cirrus on the first night, but
conditions were photometric for the other two.  The centering of the CCD frame
was off by several arc minutes east for five of the observed fields
(10,12,14,16,18).  While this introduced a few gaps in our survey
region, it did not have any impact on the presented science.

During our second run, 1995 Mar 24--28, we used the PFCCD to observe
fifteen more halo fields.  Three of the 1993 fields (4,5,21) were
re-observed because the seeing and weather were better.  In 1995, the 
PFCCD covered a $14\farcm7\times14\farcm7$ field on the sky at a
pixel scale of $0\farcs431 \ \perpix$.  Seeing was in the range
$1\farcs1$--$1\farcs8$, and weather conditions were non-photometric on
the first night (thin cirrus), but photometric for the remainder of the
run.

For both PFCCD imaging runs,
we used a narrowband filter with a central wavelength of 5016\AA\ and a
transmission FWHM of 35\AA\ in an f/2.7 beam.  
The off-band filter for these runs was a 200\AA\ wide filter centered at
5280\AA. For each field, the total exposure times were typically
2400--3600~s through the on-band filter, and 500s through the
off-band filter.  The off-band images are deeper in the continuum
to help avoid spurious PN detections.

On a third observing run, we used the CTIO Mosaic II CCD camera to image 
two major axis fields on opposite sides of the galaxy centered at
projected radii of 75 (NE) and 85 (SW) kpc.  This imaging was done on
2000 June 4 under photometric, moonless conditions with a seeing of
$1\farcs1$.  For our Mosaic imaging, we used
an on-band filter with a central wavelength of 5016\AA\ and a FWHM of
45\AA\ in an f/2.87 beam.  
In place of a true off-band filter, we substituted a Johnson {\it V} filter.  
Total exposure times were 3600s in [\ion{O}{3}] and 300s in {\it V}.

\subsubsection{PNe Detection and Astrometry}
We used standard CCD reduction software in IRAF\footnote{IRAF is
distributed by the National Optical Astronomy Observatories, which are
operated by the Association of Universities for Research in Astronomy,
Inc., under cooperative agreement with the National Science Foundation.}
to bias subtract, flat field, and remove cosmic rays from our imaging data.  
PN identification can be done in a few different ways.  The traditional
method of ``blinking'' on- and off-band images was used for some of our
survey.  We also divided the on-band by the off-band to look for bright,
stellar residuals.  Another technique, used on f14, f15, and
our Mosaic fields, is to employ an object detection algorithm such as
SExtractor (Bertin 1996), and select objects that have large color.
This is similar to the method described in Arnaboldi \etal
(2002).  Even when automated detection was employed, all PN candidates
were visually inspected.

Deriving accurate positions for PNe candidates 
is important for fiber spectroscopy,
where fibers are only 2--3$\arcsec$\ in diameter.  The fact that our
data were taken over a period of many years during which astrometric
reference catalogs continually improved compels us to be very careful
when creating an internally consistent astrometric catalog.  All
positions published in this paper are on the astrometric reference frame
of the USNO-A2.0 catalog (Monet \etal 1998).  The reference catalogs
used for spectroscopic follow-up, however, also included the 
Guide Star Catalog v1.0 (GSC; Lasker \etal 1990), GSC v1.2, and USNO-A1.0.  

Compared to the GSC v1.0 and v1.2, the USNO-A2.0 catalog has both a
higher stellar density and astrometry that is less plate-dependent. 
With NGC 5128's relatively low galactic latitude ($+19\degr$), and the 
depth of the USNO-A2.0 catalog, there was no shortage of astrometric
reference stars with which to obtain a solution---hundreds of 
stars with $B$ magnitudes between 13.5 and 17.0 were used for each field.
Typically, fourth-order polynomials were necessary to fit the data,
obtaining positional accuracy with
$< 0\farcs3$ residuals.  We used overlap between neighboring 
PFCCD fields to confirm that our astrometry was consistent
between fields.

Astrometry for the H93b sample of PNe was based on the GSC v1.0, 
which possesses a small, but
non-negligible zeropoint shift when compared to the USNO-A2.0 system.
Although the internal relative astrometry for the previously published
and the current work has RMS errors of only
$0\farcs10$--$0\farcs33$, the zeropoint shifts between systems could be
as large as $2\arcsec$.  
Our goal was to create a consistent set of positions that future
observers could observe in the same fiber configuration, and also to match
duplicate observations of the same object.  Since the relative astrometry
used for each observing run was shown to be satisfactory by proof of
successful spectroscopic observations, when making the merged, final catalog, 
we decided that simple zeropoint
shifts would suffice to make the PN positions consistent.

To calculate the appropriate shifts in the vicinity of \cena, we found
the USNO-A2.0 positions for the bright GSC reference stars listed in H93b
(their Table~2).  From these, we calculated shifts of $+0\farcs55$ in
right ascension and $+1\farcs30$ in declination (to be applied to the
H93b positions).  In all cases, we checked our new astrometry against
the PN positions in the actual images and found that they agreed.
We include in Table~\ref{table:astromrefs} a list of bright, isolated,
astrometric fiducial stars from the USNO-A2.0 catalog.  These stars
were used for guide fibers when taking spectra of PNe.  Not only can
they be used in future observations, but they can also be used to
translate the coordinates published in this paper to future astrometric
reference frames.

\begin{deluxetable}{llll}
\tablewidth{0pt}
\tabletypesize{\scriptsize}
\tablecaption{Astrometric Fiducial Stars from USNO-A2.0 
              \label{table:astromrefs}}
\tablehead{
\colhead{Name} & \colhead{RA(J2000)} & \colhead{Dec(J2000)} &
\colhead{{\it R} mag}
}
\startdata
F01 & 13:24:49.38 & $-$43:05:07.7 & 14.0 \\
F02 & 13:25:56.18 & $-$43:09:60.0 & 14.0 \\
F03 & 13:26:17.59 & $-$42:55:36.9 & 14.0 \\
F04 & 13:26:05.46 & $-$42:51:37.9 & 14.0 \\
F05 & 13:24:37.28 & $-$42:47:05.0 & 14.0 \\
F06 & 13:23:55.81 & $-$42:55:42.0 & 14.0 \\
F07 & 13:23:51.18 & $-$43:03:53.8 & 14.0 \\
F08 & 13:25:07.04 & $-$42:41:57.3 & 14.0 \\
F09 & 13:24:31.48 & $-$43:18:29.7 & 14.0 \\
F10 & 13:26:24.10 & $-$43:19:03.2 & 14.0 \\
F11 & 13:24:46.97 & $-$43:20:29.8 & 14.0 \\
F12 & 13:24:05.15 & $-$42:46:52.8 & 14.0 \\
F13 & 13:27:21.77 & $-$42:51:50.5 & 14.0 \\
F14 & 13:24:08.86 & $-$42:42:49.9 & 14.0 \\
F15 & 13:26:52.11 & $-$43:19:40.9 & 14.0 \\
F16 & 13:27:39.21 & $-$42:59:41.7 & 14.0 \\
F17 & 13:23:25.38 & $-$42:51:50.0 & 14.0 \\
F18 & 13:23:38.87 & $-$43:18:50.3 & 14.0 \\
F19 & 13:24:34.99 & $-$43:27:46.6 & 14.0 \\
F20 & 13:24:03.98 & $-$43:26:10.1 & 14.0 \\
F21 & 13:24:21.90 & $-$43:28:31.7 & 14.0 \\
F22 & 13:27:59.96 & $-$42:49:59.5 & 14.0 \\
F23 & 13:25:13.90 & $-$43:34:22.8 & 14.0 \\
F24 & 13:23:56.25 & $-$43:33:28.9 & 14.0 \\
F25 & 13:28:33.83 & $-$43:20:29.8 & 14.0 \\
F26 & 13:28:19.55 & $-$42:36:01.6 & 14.0 \\
F27 & 13:25:53.61 & $-$43:42:28.6 & 14.0 \\
F28 & 13:28:42.28 & $-$43:23:57.6 & 14.0 \\
F29 & 13:22:08.23 & $-$43:22:50.7 & 14.0 \\
F30 & 13:29:11.28 & $-$43:12:44.8 & 14.0 \\
F31 & 13:28:16.53 & $-$43:30:24.1 & 14.0 \\
F32 & 13:22:52.84 & $-$43:34:01.3 & 14.0 \\
F33 & 13:30:09.13 & $-$42:47:32.2 & 14.0 \\
F34 & 13:30:13.34 & $-$42:50:33.4 & 14.0 \\
F35 & 13:24:16.08 & $-$43:53:26.6 & 14.0 \\
F36 & 13:29:15.79 & $-$43:35:50.9 & 14.0 \\
F37 & 13:26:05.39 & $-$42:04:21.7 & 14.1 \\
F38 & 13:26:02.57 & $-$42:10:28.4 & 14.1 \\
F39 & 13:26:35.91 & $-$42:07:51.1 & 14.1 \\
F40 & 13:24:46.80 & $-$42:12:52.6 & 14.1 \\
F41 & 13:25:01.77 & $-$42:15:39.7 & 14.1 \\
\enddata
\tablecomments{Positions and magnitudes are from the USNO-A2.0 catalog.
All stars are within 1\degr\ of \cena.}
\end{deluxetable}

\begin{deluxetable*}{llcclcl}
\tablewidth{0pt}
\tabletypesize{\footnotesize}
\tablecaption{Observing Runs: Spectroscopy \label{table:obsrun_spec}}
\tablehead{
\colhead{Target(s)} & \colhead{Instrument} & 
\colhead{FOV\tablenotemark{a}} & \colhead{Fibers\tablenotemark{b}} &
\colhead{Grating} & \colhead{Res\tablenotemark{c}} & \colhead{Date} \\
\colhead{} & \colhead{} & \colhead{(\arcmin)} & \colhead{} &
\colhead{} & \colhead{(\AA)} & \colhead{}
}
\startdata
PN & AAO/2dF & 120 & 200 & 
     1200V & 2.2 & 1997 Mar 14,16 / Apr 3--5 \\
PN & CTIO/ARGUS & 50 & 24 & KPGL3 & 3.4 & 1997 May 13--14\\
PN/GC & AAO/2dF & 120 & 400 & 1200B & 2.2 & 2001 Jan 20 \\
PN/GC & CTIO/Hydra & 40 & 130 & KPGL3 & 4.9 & 2001 Feb 18--20 \\
PN/GC & AAO/2dF & 120 & 400 & 1200V & 2.2 & 2002 Apr 4--6, 9--10 \\
PN (H$\alpha$) & AAO/2dF & 120 & 400 & 1200V & 2.2 & 2002 Jul 7 \\
\enddata
\tablenotetext{a}{Field-of-View}
\tablenotetext{b}{Number of fibers available for each configuration}
\tablenotetext{c}{Spectral Resolution}
\end{deluxetable*}

Although these data were taken under conditions of variable transparency
and seeing, and we did not apply photometric calibrations to them,
the limiting magnitude of each image in this new survey is at least as
faint or fainter than that of H93a, which was complete to $m_{5007} =
24.8$, or 1.5 magnitudes down the PN luminosity function.  
In all cases where our
fields overlapped area previously surveyed, we recovered all of the PNe
listed in H93b.  In addition, we were able to identify additional
candidates in these regions that were not detected in the previous study.
In total, we identified 356 new PN candidates which, in addition to the
785 PNe listed in H93b, brings the total number of [\OIII] imaging
detections in \cena\ to 1141.  Of these 1141, 431 PNe were previously
confirmed spectroscopically by H95.  In the following section, we
describe our effort to obtain radial velocities for the remaining
candidates.

\subsection{Spectroscopy}
\subsubsection{Observations}
\label{sec:pnobs_spec}

Spectroscopy was done over the course of six separate observing runs:
four with the Anglo-Australian Observatory (AAO) 2-degree Field (2dF) 
fiber spectrograph, one with the CTIO Argus spectrograph, and one with
the CTIO Hydra spectrograph.  The 2dF spectrograph in particular is an ideal
instrument for \cena\ studies because its field-of-view encompasses a
60~kpc radius around the galaxy center.  Central crowding of PNe is the
primary limitation that prevents maximal fiber allocation.  
We targeted the 356 new candidates from
the 1993--2000 imaging data as well as 353 candidates from the
sample of H93b
that had no previous velocity measurement.  On three of these
runs, fiber configurations also included globular cluster candidates
in \cena\ (Peng, Ford, \& Freeman 2004b).
For the first five runs, the spectral range was centered on
the [\ion{O}{3}]$\lambda5007$ line, and for the last run it was centered on
H$\alpha$.  The spectrograph setups, fields of view, and resolutions
are presented in Table~\ref{table:obsrun_spec}.

Our first AAO/2dF observing run occurred during parts of four nights in the
spring of 1997.  This period was designated as ``shared-risk'' for 2dF,
and only one spectrograph (with 200 fibers) was available.  The weather was
clear and seeing was $1\farcs0$--$1\farcs5$ for all three nights.  
Total exposure time for each configuration ranged from 1--2.5 hours.

In May 1997, we used the ARGUS multi-fiber spectrograph on the CTIO
4-meter telescope to observe PNe candidates from the 1995 imaging data, as
well as some from the 1993 data set.  At the time of the observations, ARGUS
contained 24 fiber pairs, each $1\farcs8$ in diameter.  
We observed seven fiber configurations over two nights, with each
configuration receiving 2.25~hours of exposure time.

Our second AAO/2dF observing run was taken in service mode in January,
2001.  We received one 2.5~hour exposure for a single fiber configuration.  
This is the first of three runs where we also placed fibers on globular
cluster candidates.

In February, 2001, we observed with the Hydra multi-fiber spectrograph on
the CTIO 4-meter.  We obtained data for five fiber configurations over
three nights of clear weather, and $1\farcs2$--$2\farcs0$ seeing.  
The spectral resolution of this data is lower than for other runs, but
this did not seem to affect the velocity accuracy.

Our third 2dF run was in April, 2002 during bright conditions. 
We used a total exposure time of 7.5~hours for one fiber configuration over
the course of five nights.

Our fourth 2dF data set was taken in service mode during July, 2002.
This one differed from previous runs in that the grating was centered on
the H$\alpha$ emission line.  Previously unobserved PNe that were
targeted in this fiber setup were from fields 14 and 15.  

\subsubsection{Data Reduction}

With the development of a number of dedicated software packages, the
reduction of fiber spectroscopy is relatively routine.  For our
reductions we used either the 2dF data reduction package ({\it 2dfdr}), or the
IRAF packages {\it doargus} and {\it dofibers}.  At the time of our first 2dF
observations, {\it 2dfdr} was still being developed, so we chose to reduce our
data with {\it dofibers} in IRAF.

In a typical reduction, images were bias subtracted and the spectra were
extracted using a fiber flat field exposure.  The extracted spectra were
then flat-fielded and wavelength calibrated.
Arc lamp exposures, taken through the night, showed that the
wavelength solution did not change appreciably over time.  Nevertheless, 
we used the nearest arc exposure (in time) to dispersion correct each
exposure before combining.  RMS residuals in the
wavelength solution were typically $\sim0.15$\AA.
The spectra were throughput corrected using either offset sky exposures
(2dF), sky line fluxes (2dF/H$\alpha$), or projector flats (Hydra).
A median sky was created using 20--40 sky fibers, and then was
subtracted from each spectrum.
Finally, the multiple exposures from each fiber configuration were
regridded to a common dispersion solution and averaged with a 3-$\sigma$
rejection threshold for cosmic rays.

There were some exceptions to this description of the reduction process.
For the Argus data and first 2dF data, flat fielding, throughput
corrections, and sky
subtraction were not applied---these calibrations do not affect the
measurement of emission line radial velocities.  Also, 
the fiber flats taken during our first 2dF run were not appropriate for
aperture extraction, so the arc exposures were used instead. 

\vspace{1cm}
\subsubsection{Velocity Measurement}

We visually inspected each combined spectrum for the presence of the
[\ion{O}{3}]$\lambda5007$ emission line.  
For those spectra where there was discernible [\ion{O}{3}] emission,
we fit the line centers using the IRAF package {\it rv}, and applied
heliocentric corrections to the measured doppler velocities.
We noted the detection of [\ion{O}{3}]$\lambda4959$ in bright PNe, but
did not include this line in the velocity measurement.

We also verified that the lines were symmetric and unresolved.  
Previous surveys for intracluster PNe in Virgo 
have accidentally detected \lya
emitting galaxies at high redshift (Kudritzki \etal 2000).  
In these galaxies, the
$\lambda1216$ line is redshifted into the on-band filter.  For the
filters we used to survey \cena, a \lya galaxy would need to be
near $z=3.125$ to be detected.
The detection of both $\lambda\lambda4959,5007$ lines
confirms that the observed lines are from [\ion{O}{3}].  For those
spectra where only one line is detected, we can check for 
an overly broad or asymmetric line profile, a signature of \lya.  
All one-line PN spectra were checked manually, and none were obvious
\lya galaxies.  This is not surprising because \cena\ is over three
times closer than the Virgo cluster.  Our images can be shallower than
those for surveys in Virgo and still detect PNe that are 2.5 magnitudes
down the PN luminosity function (similar to H93a).  
By re-observing PNe with published photometry (H93b), we
find that the faintest objects for which we measure velocities have
$F_{5007}\sim2\times10^{-16}\ {\rm erg\ s^{-1}\ cm^{-2}}$, which is
approximately two magnitudes down the PN luminosity function.
By comparison, in a narrowband blank-field survey to measure the surface
density of background galaxy contaminants, Ciardullo \etal (2002) found
zero galaxies that had line fluxes this bright, albeit in a
significantly smaller area of sky.  This is consistent with early
surveys for $z>3$ emission line galaxies (summarized by Prichett 1994) 
which were unsuccessful and had depths similar to or deeper than
our PN survey.  As a result, high redshift \lya galaxy
interlopers are not of a problem for this PN survey of \cena.

\section{The Planetary Nebula Catalogs}
\label{sec:pncat}

Ultimately, we wished to create a table of all the unique PNe in 
\cena\ with their positions and best radial velocities.  This task was
complicated by imaging overlap regions, two astrometric systems and
multiple observing runs.  Below, we outline the steps we took to produce
the final catalogs.

\subsection{Duplication and Velocity Accuracy}

Once our entire catalog was on a common astrometric system, we matched
each PN against the entire catalog to check duplication.  Mainly, duplication
occurred in overlap regions where PNe were identified in separate images
and given different names.  In cases where a ``new'' PN was actually
previously observed in H95, we assigned a new composite name.  When two
PNe were both previously unpublished, one name was chosen.  In the
process of checking for duplicates, we discovered that one PN in the
H95 catalog, 5504, was observed twice and listed twice.  Another pair
of PNe, 5420 and 5423, were found to be the same object.
In both cases, we averaged
the two velocity measurements and removed one listing from our catalog.
The correct number of unique PNe in H95 should be 431.

Besides the unintentional duplication, we
also purposely repeated spectroscopic observations of certain PNe over
multiple observing runs.  Together, these PNe allowed us to 
quantify systematic velocity offsets and random errors.
For consistency, the ``zeropoint''
velocity reference frame is considered to be the one defined by the PNe 
published in H95.  All
data from subsequent observing runs were tied back to the H95
sample by using a sample of re-measured PNe.

Systematic velocity shifts and empirical velocity errors derived from
these duplicate and multiply observed PNe are presented in
Table~{\ref{table:duplicates}.
The largest systematic shift between any two of the newly presented
spectroscopic runs is 11~\kms, while the largest shift between one of
the new spectroscopic runs and H95 is 21~\kms.  
All of the velocity offsets between our newer data
and H95 have the same sign, and may point to some systematic effect.
However, given the small size of the offsets, 
we chose to bring our velocities onto
the system established by H95 for the sake of consistency in the literature.
Multiple velocity measurements of the same PN were averaged after
applying this offset.  Because some PNe from H95 were re-observed, we
list them in our catalog with a composite name and their new velocity.

\begin{deluxetable}{lccc}
\tablewidth{0pt}
\tabletypesize{\footnotesize}
\tablecaption{Velocity Consistency and Errors\label{table:duplicates}}
\tablehead{
\colhead{ObsRun\tablenotemark{a}} & 
\colhead{${\Delta}v$\tablenotemark{b}} & 
\colhead{$\sigma/\sqrt{2}$} & 
\colhead{PN\tablenotemark{c}} \\
\colhead{} & \colhead{(\kms)} & \colhead{(\kms)} & \colhead{}
}
\startdata
2dF97 & $-21$ & 15.9 & 51 \\
Argus97\tablenotemark{d} & $-19$ & 15.6 & 17 \\
2dF01 & $-17$ & 14.0 & 16 \\
Hydra01 & $-10$ & 28.3 & 11 \\
2dF02 & $-20$ & 11.2 & 15 \\
2dF02(H$\alpha$) & $-13$ & 19.2 & 152 \\ 
\enddata
\tablenotetext{a}{Refers to observing runs listed in
Table~\ref{table:obsrun_spec}}
\tablenotetext{b}{Add ${\Delta}v$ to match H95 velocity frame}
\tablenotetext{c}{Number of PNe in common with H95}
\tablenotetext{d}{The Argus97 sample had no PNe in common with H95.  
These values are based on 17 PNe in common with the 2dF97.}
\end{deluxetable}

These repeat observations also allow us to determine the internal
velocity measurement errors.  We estimate the RMS error
as the standard deviation about the mean velocity difference, divided by
$\sqrt{2}$ (making the
approximation that the errors are roughly equal for each observing run).  
RMS velocity errors were typically $\sim20$~\kms, which is comparable to
the best errors obtained by H95.
This is consistent with the velocity error we derived from our first 2dF
run, where 11 PNe were observed in multiple fiber configurations.
This internal comparison yielded a velocity RMS error of $10$~\kms.
One caveat to this method is that the PNe with duplicate observations
tend to be ones with stronger $\lambda5007$, so these errors, while not
atypical, should be considered lower limits. 

\subsection{The Catalogs}

\begin{deluxetable*}{lcccrrr}
\tablewidth{0pt}
\tabletypesize{\scriptsize}
\tablecaption{\cena\ Planetary Nebula Position and Velocity Catalog}
\tablehead{
\colhead{ID} & \colhead{V} & 
\colhead{RA(J2000)} & \colhead{Dec(J2000)} &
\colhead{X} & \colhead{Y} & \colhead{R} \\
\colhead{} & \colhead{(\kms)} & 
\colhead{(hh:mm:ss.ss)} & \colhead{(dd:mm:ss.s)} &
\colhead{(\arcmin)} & \colhead{(\arcmin)} & \colhead{(\arcmin)} \\
\colhead{(1)} & \colhead{(2)} & 
\colhead{(3)} & \colhead{(4)} &
\colhead{(5)} & \colhead{(6)} & \colhead{(7)}
}
\startdata
Ctr003      & 366 & 13:25:48.88 & $-$42:56:54.5 &   5.650 &  $-$0.771 &   5.684 \\
ctr007      & 483 & 13:25:43.77 & $-$42:55:03.8 &   6.626 &   1.052 &   6.676 \\
ctr010      & 370 & 13:25:42.29 & $-$42:55:44.1 &   5.920 &   0.890 &   5.946 \\
ctr013      & 233 & 13:25:40.89 & $-$42:55:16.7 &   6.148 &   1.361 &   6.248 \\
ctr026      & 521 & 13:25:44.86 & $-$42:57:44.8 &   4.541 &  $-$0.649 &   4.617 \\
ctr107      & 443 & 13:25:34.50 & $-$42:58:27.1 &   2.877 &   0.500 &   2.872 \\
ctr109      & 542 & 13:25:24.86 & $-$43:02:37.6 &  $-$1.554 &  $-$0.450 &   1.555 \\
ctr115      & 270 & 13:26:19.48 & $-$42:56:53.9 &   8.861 &  $-$5.358 &  10.370 \\
ctr118a     & 251 & 13:25:36.61 & $-$42:58:20.8 &   3.184 &   0.244 &   3.209 \\
ctr118b     & 573 & 13:25:27.21 & $-$43:09:39.4 &  $-$7.066 &  $-$4.834 &   8.530 \\
\enddata
\label{table:pntable}
\tablecomments{There are 780 confirmed PNe.  
The complete version of this table is in the electronic
edition of the Journal.  The printed edition contains only a sample.}
\end{deluxetable*}

After removing duplicate measurements and incorporating systematic velocity
shifts, we merged the samples to yield a catalog of 780
spectroscopically confirmed PNe.  Of these,
431 are from H95, and 349 are new.  Also, 148 PNe are at projected
radii greater than 20~kpc.  The names, positions, and velocities of
all PNe are presented in Table~\ref{table:pntable}.
The columns are: (1) PN identification (field and PN number,
H93b designation, or composite name),
(2) radial velocity in \kms, (3) RA (J2000), (4) Dec (J2000), 
(5) distance along photometric major axis ($X$) in arc minutes, 
positive to the northeast as in H95, (6) distance along photometric
minor axis ($Y$) in arc minutes, positive to the northwest,
and (7) projected radius from the galaxy center in arc minutes.

The positional quantities $X$ and $Y$ follow the convention of
H95.  The position angle of the major axis is $35\degr$,
as measured in Dufour \etal (1979).  Because our survey area is over $2\degr$
across, the small angle approximation for distances on a sphere can be
incorrect by as much as $\sim10\arcsec$.  Instead, we transformed the
equatorial coordinates to a spherical system with the origin at the
center of \cena\ and the axes defined by the photometric axes.  
We adopt the center of \cena\ to be the value given by Kunkel \& Bradt
(1971), which is ($13^{\rm h}25^{\rm m}27\fs72$, 
$-43\degr01\arcmin05\farcs8$, J2000).  This is
the same value used in H95.  The pole
of this system is at ($09^{\rm h}07^{\rm m}35\fs9$, 
$+24^{\rm d}47^{\rm m}39\farcs63$, J2000), and the
longitude of the north celestial pole in this system is
$138\fdg72103$.  The transformed
values are given with enough precision to recover the original
equatorial coordinates.

We also present a catalog of PN candidates detected in our imaging for
which we do not have a velocity because they have not been observed
spectroscopically.  There are 361 remaining PN candidates, and they 
are listed in Table~\ref{table:notobs}.  

\begin{deluxetable}{lcccrrr}
\tablewidth{0pt}
\tabletypesize{\scriptsize}
\tablecaption{Unconfirmed PN Candidates in \cena}
\tablehead{
\colhead{ID} & \colhead{RA(J2000)} & \colhead{Dec(J2000)}
}
\startdata
     212 & 13:25:36.95 & $-$43:00:59.6 \\
     215 & 13:25:39.09 & $-$43:01:01.9 \\
     227 & 13:25:35.18 & $-$43:00:31.0 \\
     228 & 13:25:36.67 & $-$43:00:11.5 \\
     229 & 13:25:41.42 & $-$43:01:25.2 \\
     231 & 13:25:33.86 & $-$43:00:57.4 \\
     232 & 13:25:35.00 & $-$43:00:13.0 \\
     235 & 13:25:52.98 & $-$43:00:55.8 \\
     238 & 13:25:49.47 & $-$43:01:43.7 \\
     239 & 13:25:35.68 & $-$43:00:41.2 \\
\enddata
\label{table:notobs}
\tablecomments{There are 361 unconfirmed PNe. 
The complete version of this table is in the electronic
edition of the Journal.  The printed edition contains only a sample.}
\end{deluxetable}

\subsection{X-ray matches}

We compared our PN catalog with the catalog of X-ray point sources
in \cena\ presented by Kraft \etal (2001).  Three X-ray sources are
within 2\arcsec\ of a known PN, although one of these matches (K-053) is
more likely to be with with a neighboring continuum object.  These PN
have IDs 525, 4225, and 5203, and they match with X-ray sources K-053,
K-080, and K-241.

\vspace{1cm}

\section{Results}
\subsection{The PN Spatial Distribution}

Our catalog of PN velocities is unique in its size and spatial coverage.
We plot the positions of each confirmed PN 
on an image of \cena\ in Figure~\ref{figure:pnspatial}. 
The outline of our survey area is also drawn, and follows the
outer boundaries of the fields shown in Figure~\ref{figure:pnfields}.
There are a few gaps in our coverage within this region, mainly due to
the small CCDs used in the original H93 survey, but also due to a few
errors in pointing.  The gap between fields 8 and 10 is mostly covered
by the H93 survey, but the 6\arcmin\ gaps between fields 14/15, 16/17,
and 18/19 are real.  Besides these small regions, almost all of the area
within this boundary has been observed to a uniform depth.

\begin{figure*}
\centerline{\includegraphics[width=5.5in]{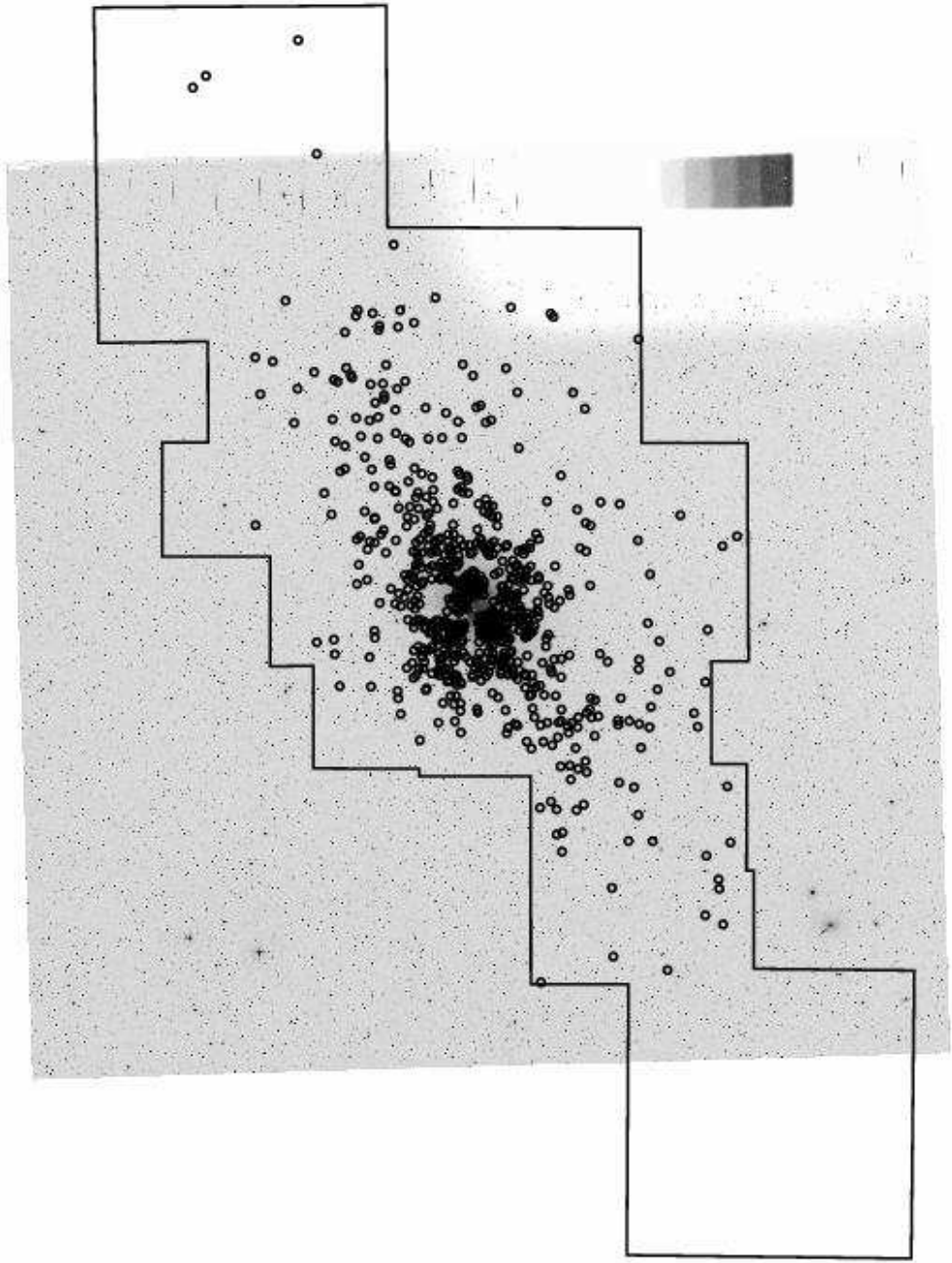}}
\caption[Spectroscopically confirmed PNe in \cena]
{Spectroscopically confirmed PNe in \cena.  
As in Figure~\ref{figure:pnfields}, this image is a $2\degr\times2\degr$
DSS image centered on \cena.  PNe are represented by circles, and the
line marks the outer boundary of our survey region.  There are a few
gaps in our sky coverage of the inner regions due to the small CCDs used
in the original H93 survey (see Figure~\ref{figure:pnfields}).
\label{figure:pnspatial}
}
\end{figure*}

\begin{figure*}
\plotone{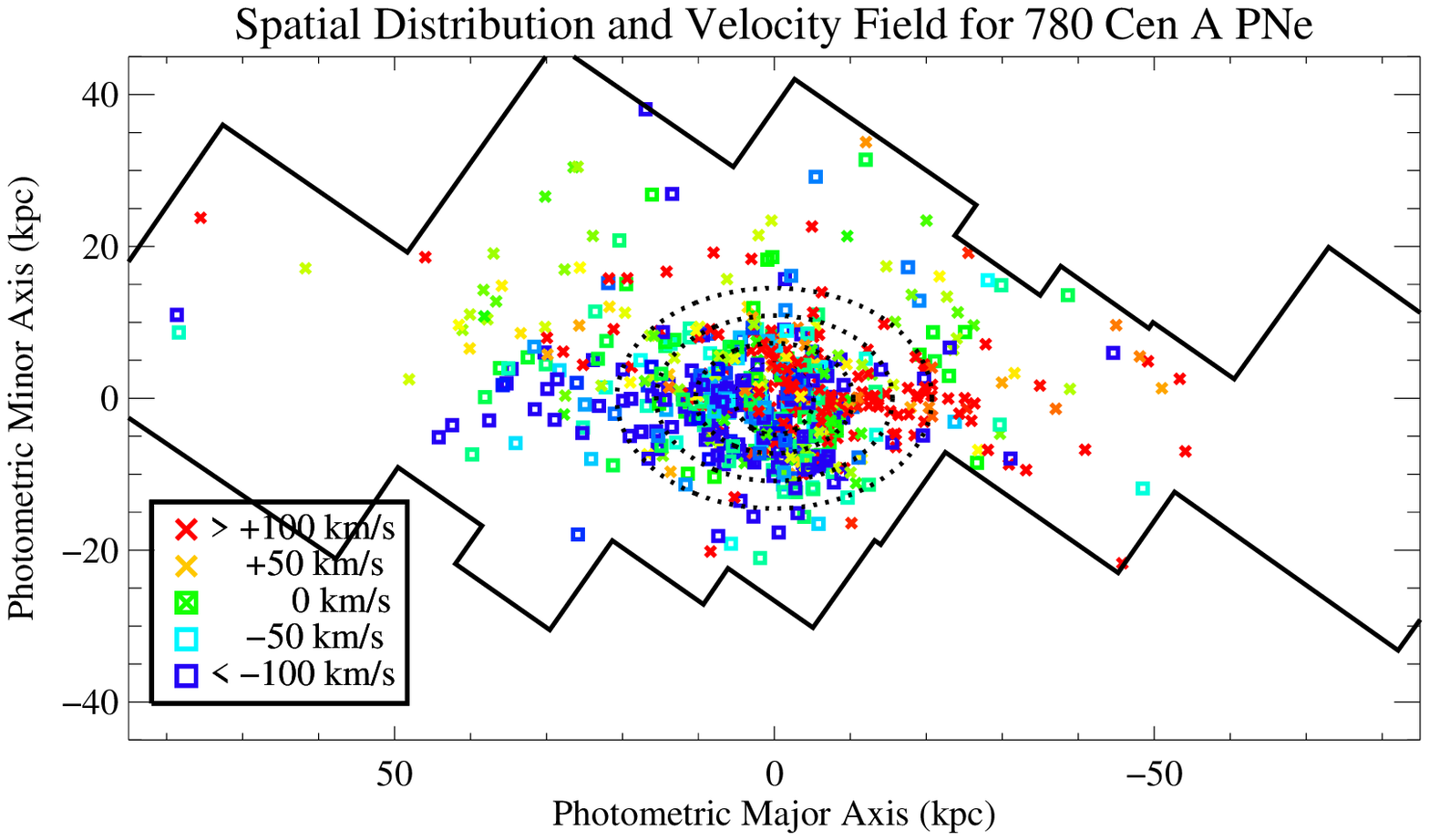}
\caption[Raw PN velocity field]
{\small Raw PN velocity field. The X-axis is along the
photometric major axis (positive is NE), and the Y-axis is along the
photometric minor axis (positive is NW).  
Dotted ellipses trace the rough isophotes for \cena\ from 1--4~$r_e$.
As in Figure~\ref{figure:pnspatial}, the solid line traces the outer
boundary of our survey area.  Each point represents the location and radial
velocity of a single PN.  PN with velocities larger than systemic
(receding) are represented by crosses, and those with velocities
smaller than systemic (approaching) are represented by open boxes.  The
color of each point shows the magnitude of the velocity with respect to
systemic.  Red points are rapidly receding, green points have velocities
near zero, and blue points are rapidly approaching.
\label{figure:rvs_pnarea}
}
\end{figure*}


Most of the PNe lie near the photometric major axis of the galaxy,
which is at a position angle of $35\degr$.  
This is especially so in the halo, where the PNe mirror 
the disky structure seen in deep photographic images (Malin 1978, also
shown in Israel 1998).
While there are three PNe at projected radii of 80~kpc in our NE Mosaic
field, there is a notable 
absence of PNe in our southwest Mosaic field.  This is not surprising,
however, as the low surface brightness halo light is known to be
somewhat asymmetric, extending more toward the northeast.
Qualitatively, the PNe do appear to follow the low surface brightness
halo light seen in Malin's image and in our own observations, and are
able to probe regions even fainter than those accessible to broadband
photometry.

The detection along the NE major axis of three PNe at $\sim80$~kpc with
normal velocities (i.e.\ they are bound to the galaxy) supports the idea
that the stellar halo extends to these distances and beyond.  Even
though these three PNe are spatially close in projection, possibly suggesting
some form of halo substructure, we think it is unlikely that they are
associated with either each other or the dwarf
irregular galaxy ESO~324-G024, which is just under 20~kpc away.
ESO~324-G024 has an optical radius of under 2~kpc, the individual PNe are
over 15~kpc apart, and they collectively have a velocity dispersion of
172~\kms.  The large differences in their radial velocities can
be more easily explained by an extension of the velocity field seen in the
inner halo.

\subsection{The PN Velocity Field}
\label{section:pnvfield}

\begin{figure*}[t]
\plotone{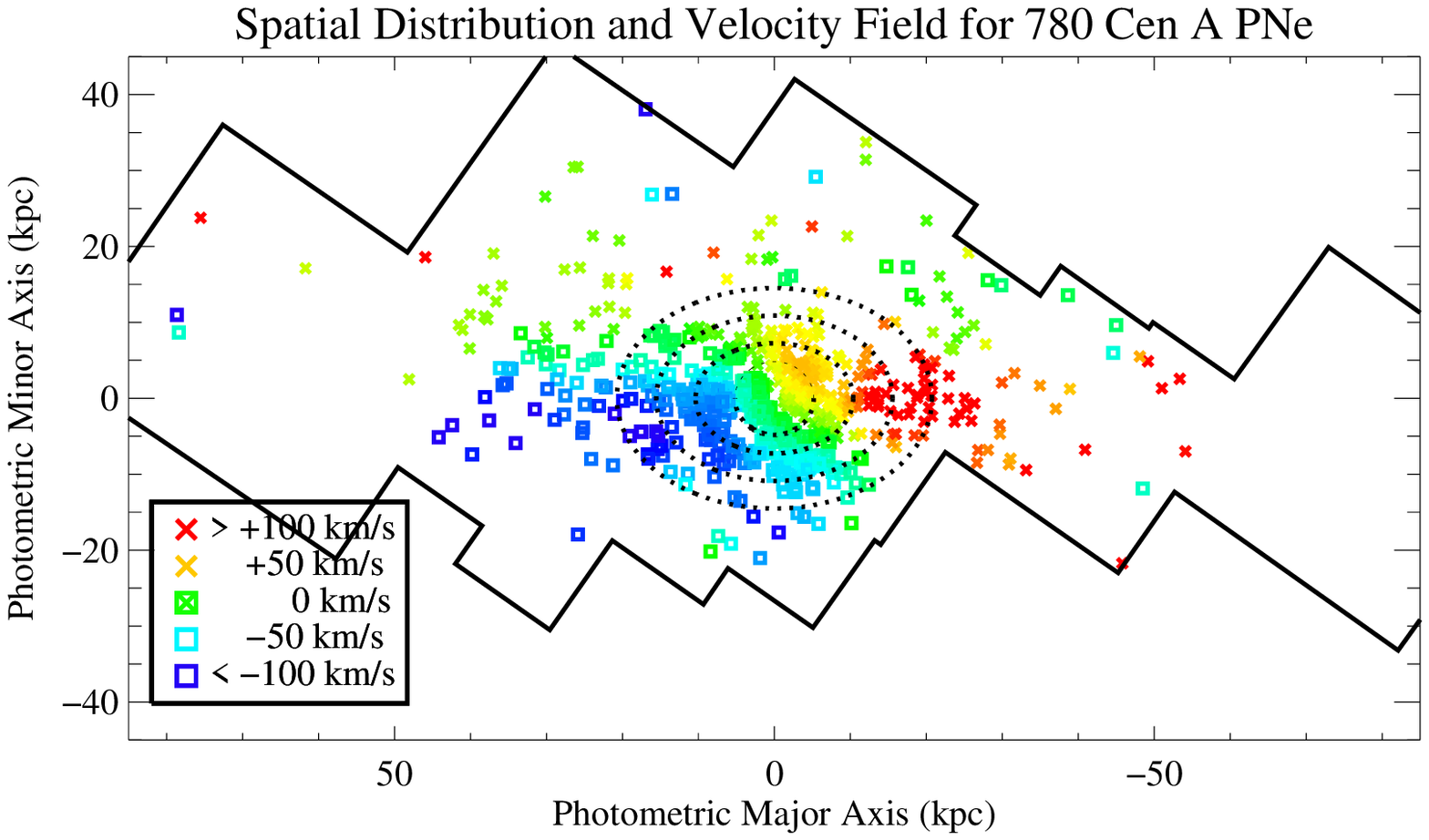}
\caption[Kernel smoothed PN velocity field]
{\small Kernel smoothed PN velocity field.  Definition of
axes and points are the
same as in Figure~\ref{figure:pnspatial}.  We estimated the mean
streaming motion at the location of each PN using a non-parametric
local linear smoother with a 3~kpc bandwidth Gaussian kernel function.
Significant rotation is visible about both the major and minor axes.
\label{figure:rvs_smooth}
}
\end{figure*}

\begin{figure*}
\plotone{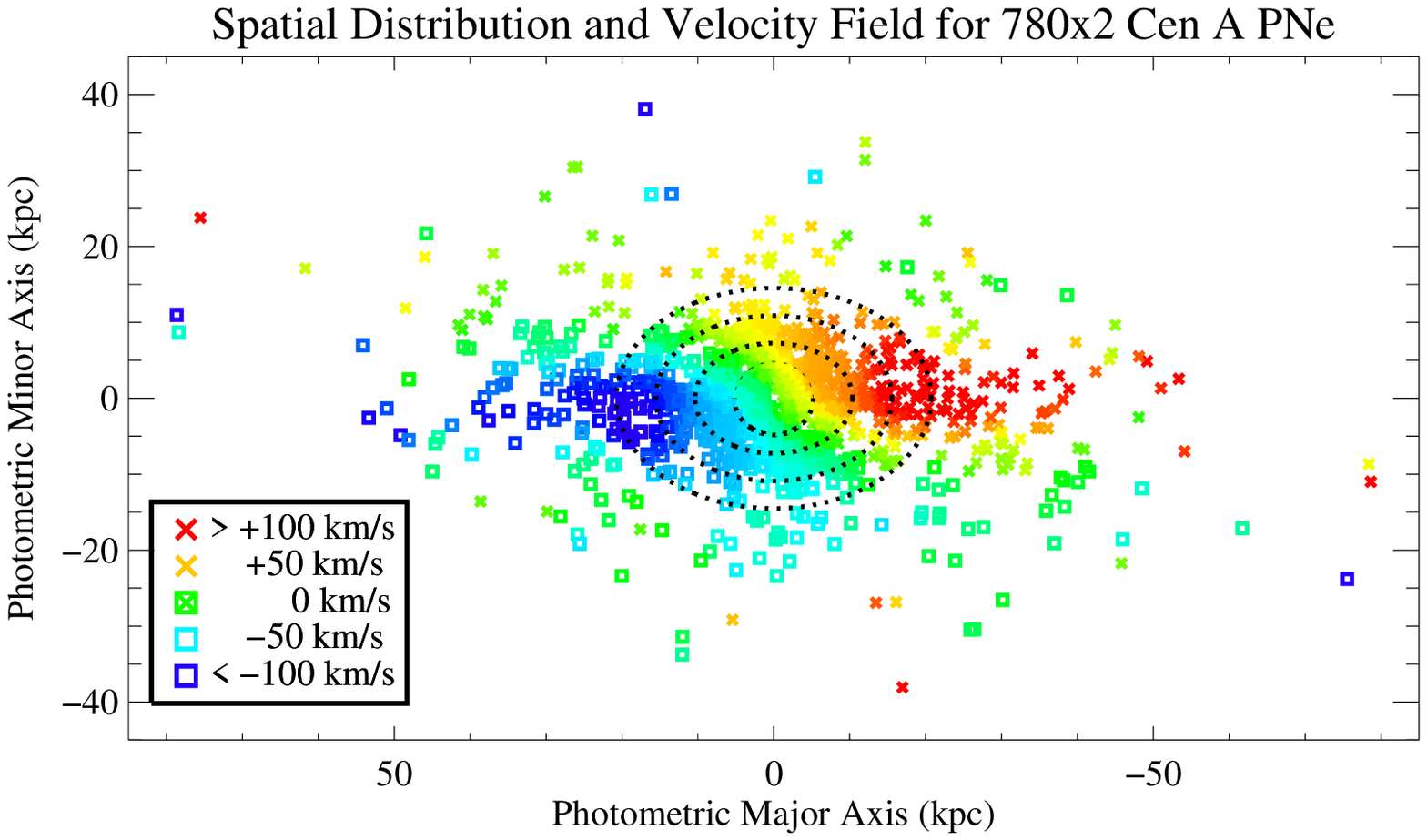}
\caption[Smoothed, antisymmetrized PN velocity field]
{\small Smoothed, antisymmetrized PN velocity field.
This figure is similar to Figure~\ref{figure:rvs_smooth} except that we
have taken advantage of the point-symmetry in the velocity field, as
expected for a triaxial potential.  In addition to the original PNe
shown in Figure~\ref{figure:pnspatial}, the position of each PN was
reflected through the origin and the sign of its velocity was reversed.
Together, these PN were kernel-smoothed as in
Figure~\ref{figure:rvs_smooth}.
\label{figure:rvs_reflect}
}
\end{figure*}

Both the density of PN velocities in the inner regions and the number of
PNe velocities in the outer halo
make our sample a valuable resource for dynamical studies.  Unlike
traditional major and minor axis long-slit spectroscopy of ellipticals
(e.g.\ Franx, Illingworth, \& Heckman 1989), discrete velocity tracers
such as PNe and globular clusters are more conducive to producing full,
two-dimensional velocity fields at large radii.  Ongoing kinematic
surveys with the planetary nebula spectrograph (Douglas \etal\ 2002), 
and integral field spectrographs like SAURON (Bacon \etal\ 2001), 
promise that two-dimensional velocity fields will be much more common in
the coming years.

In Figure~\ref{figure:rvs_pnarea}, we show our raw PN velocity field.
We plot in coordinates centered on \cena, where the X-axis is the
photometric major axis (positive is northeast), and the Y-axis is the
photometric minor axis (positive is northwest).  
Dotted ellipses trace the rough isophotes for \cena\ from 1--4~$r_e$.
These conventions are the same as those used in H95.  As in
Figure~\ref{figure:pnspatial}, the solid line traces the outer boundary
of our survey area.  Each point represents the location and radial
velocity of a single PN.  PNe with velocities larger than systemic
(receding) are represented by crosses, and those with velocities
smaller than systemic (approaching) are represented by open boxes.  The
color of each point shows the magnitude of the velocity with respect to
systemic.  Despite the large velocity dispersion, the bulk rotation
along the major axis described by H95 clearly extends farther into the halo.

Previous work with discrete velocity tracers, such as PNe or globular
clusters, use a variety of techniques to extract bulk kinematics.
The size of the sample usually dictate the methods used.  For small
samples, it is often beneficial to assume a reasonable parametric 
model, such as
solid-body rotation or a flat rotation curve.  In cases where there are
sufficient numbers of test particles, some have 
used non-parametric methods to avoid {\it a priori} assumptions of the
form of the mean velocity field.  For example, Merritt, Meylan \& Mayor
(1997) used the method of thin plate smoothing splines 
in conjunction with generalized cross-validation and penalized
likelihood techniques to recover the velocity distribution of stars in
the Galactic globular cluster $\omega$~Centauri.  
In NGC~1316, Arnaboldi \etal (1998) used a similar technique and
compared their results from those derived from parametric estimates. 
As data sets continue to grow in size, non-parametric techniques are
likely to become more common.

We chose to represent \cena's kinematics by estimating line-of-sight 
velocity moments as a function of position with a non-parametric
smoothing algorithm.  At the position of each PN, 
we applied a local linear smoother using a bivariate Gaussian kernel
function, $K_h(x^\prime,y^\prime)$, with bandwidth $h$ (which is $\sigma$
for a Gaussian kernel).  
This estimator has the advantage of being relatively simple and 
computationally fast.  Like all smoothers, it requires that we make
decisions regarding both the order of the approximation and the smoothing
scale.  In the general case of local polynomial smoothers, we need
to choose the order of the polynomial.  We used a linear smoother, 
as there was
little reason to use higher order polynomial smoothers given the high
intrinsic stellar velocity dispersion.  Next, it is important that a
smoothing scale is chosen that best matches the data.  Small bandwidths for
sparsely sampled data do not smooth over the dispersion 
in the velocity field, while
large bandwidths in rapidly varying regions may wash out real
features.  Bandwidths can also be allowed to vary as a function of local
density.  
Keeping these trade-offs in mind, we settled on a fixed bandwidth of
$h=3$~kpc for our
kernel.  This smoothing scale was the smallest one that did not preserve
noisy features in densely sampled regions.
We also found that fixing the bandwidth did not result in a loss of
information in the outer halo, and has the benefit of making
the velocity field easier for readers to interpret.  The final results
are not strongly sensitive to the bandwidth size. However, for
$h>5$~kpc, we lose resolution within 1~$r_e$, a region
where the velocity field shows interesting structure.

The resulting velocity field is shown in Figure~\ref{figure:rvs_smooth}, where
each PN's velocity has been replaced with $\hat{v}(x,y)$,
the estimate of the mean velocity field at that position.  This smoothed
velocity field shows much more clearly the bulk motion
of the stars in \cena.  One can see the fast, disk-like rotation of
stars along the major axis, as well as significant minor axis rotation.

Figure~\ref{figure:rvs_smooth} also shows point symmetry in the velocity
field --- i.e.\ stars on opposite sides of the origin have opposite
velocities with respect to the galaxy, $(x,y,v)\rightarrow(-x,-y,-v)$.
This symmetry is expected for
triaxial potentials, and can be exploited to double the number of data
points.  We tested the validity of assuming point symmetry by
calculating the correlation between the original and antisymmetric
(reflected through the origin) velocity fields.  We computed the
Spearman rank-order correlation coefficient, which has a range of $-1$
for perfectly anti-correlated samples to $+1$ for perfectly correlated
samples.  Restricting ourselves to a region of \cena\ that is
well-sampled, a $100\times30$~kpc box with 724~PNe, we compute a
correlation coefficient of 0.84.  This rejects by $23\sigma$ the null
hypothesis---that the reflected field is uncorrelated with the original.

Because of this symmetry, we can create a velocity field with 1560
velocities.  This field contains both the original PNe as well as their
reflected counterparts.  The smoothed, antisymmetrized velocity field is
shown in Figure~\ref{figure:rvs_reflect}.  Two things are immediately
apparent in this 
velocity field.  First, the strong major axis rotation is reinforced,
extending to at least 50~kpc, and maybe as far out as
80~kpc.  There is also weaker but still significant rotation along the
minor axis that extends as far as the data goes.
Second, the line of zero-velocity is both misaligned and twisted with
respect to the photometric axes.  
Both of these effects are expected when observing a triaxial system from
certain viewing angles (Statler 1991).

The twisting of the zero velocity contour (ZVC) is more evident in
Figure~\ref{figure:rvs_zerov}, where we display only those positions
that have velocities within $\pm12$~\kms\ of zero.  At the center of the
galaxy, the ZVC is almost parallel to the minor axis.  At a distance
of 1~$r_e$, however, it turns and follows a straight line at an angle of
$7\degr$ from the major axis.  We parametrize the ZVC as
\[ y = \frac{x \; (0.125 \, |x|+5)}{\sqrt{x^2 + 2.56}} \]
where $x$ and $y$ are in kpc.  We then plot this curve on the full
antisymmetrized velocity field in Figure~\ref{figure:rvs_zvc}.  Notice
how the velocity gradient is very sharp on the ZVC side of the major axis.
The rapid major axis rotation only exists in a narrow region,
within $\pm7\degr$ of the photometric major axis.  The rest of the
galaxy halo appears to be undergoing a slower minor axis rotation.  This
division even appears to be true in the outer halo at 80~kpc, where PNe
are in close proximity to each other but straddle the ZVC and thus 
have markedly different velocities.

\begin{figure}[t]
\plotone{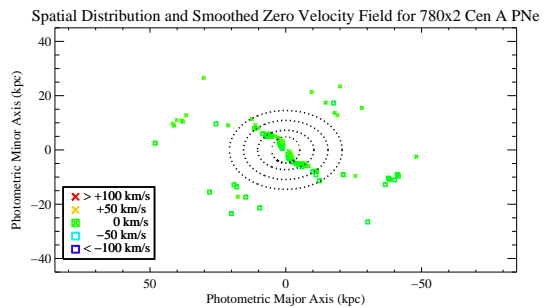}
\caption[Zero Velocity points in the \cena\ velocity field]
{\small Zero Velocity points in the \cena\ velocity field.
To better illustrate the line of zero velocity, this figure reproduces
the mean velocity field of Figure~\ref{figure:rvs_reflect}, but only
plots locations that have velocities that are within $\pm12$~\kms\ of zero.  
Notice how the zero-velocity contour starts out aligned with the minor
axis, but then has a sharp kink at about 1~$r_e$ where it proceeds
at at angle of $\sim7\degr$ to the major axis.
\label{figure:rvs_zerov}
}
\end{figure}

\begin{figure*}
\plotone{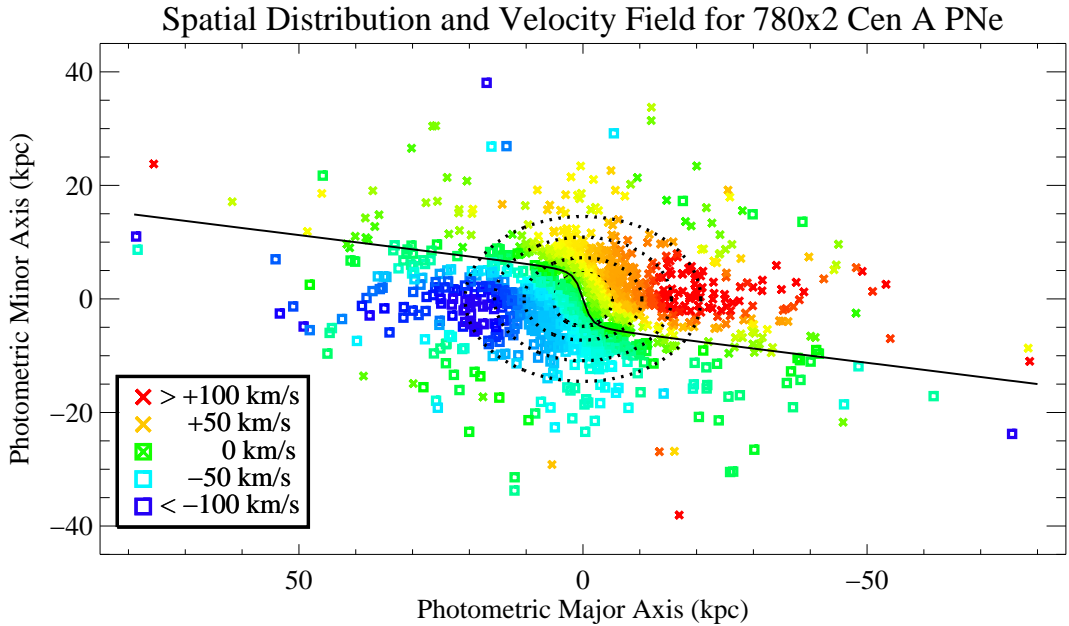}
\caption[Smoothed antisymmetrized velocity field with zero velocity
contour (ZVC)]
{\small Smoothed antisymmetrized velocity field with zero
velocity contour (ZVC) plotted.  Notice how the rapid major axis
rotation is only present within $\pm7\degr$ of the major axis, while the
rest of the galaxy halo undergoes a slower minor axis rotation. 
\label{figure:rvs_zvc}
}
\end{figure*}

\subsection{The PN Velocity Dispersion Field}

We can also calculate the two-dimensional spatial dependence of the
velocity dispersion, the second moment of the velocity distribution.
We use the same local linear smoother described above except that the
quantity in which we are interested is now $\sqrt{\hat{v^2}}$.
Calculating variance is always dependent on either model
assumptions or smoothing scales and
both parametric and non-parametric methods are prone to bias.
Napolitano \etal (2001) point out that bilinear fits to
a discrete sample of radial velocities can introduce bias in estimates
of the velocity dispersion.  This can be a problem in the sparsely sampled
outer regions where a small smoothing scale can underestimate the
dispersion, and over-smoothing can overestimate it.

In Figure~\ref{figure:rvs_vdisp}, we present our attempt at producing a
velocity dispersion field for \cena.  This figure differs from previous
ones in that we used a varying bandwidth kernel to partially compensate
for the sparse sampling in the halo.  We set the bandwidth to the larger
of 3~kpc or the distance to the tenth nearest neighbor.  While the
estimated dispersions beyond $\sim5~r_e$ are likely to be biased, we can
still see real and interesting structure in the inner halo.  The colors
in Figure~\ref{figure:rvs_vdisp} correspond to the degree of pressure
support, with red points being ``hotter''.  As one
would expect for an elliptical galaxy, the center has a high velocity
dispersion.  The velocity dispersion stays high ($\sigma\sim150$~\kms) then
falls gradually along the major axis.  Along the minor axis, the
dispersion drops more rapidly.  This agrees with the measurements of
H95, who found that the velocity dispersion changed as a function of
position angle with the highest dispersion being where the rotation was
also highest.

\begin{figure}
\plotone{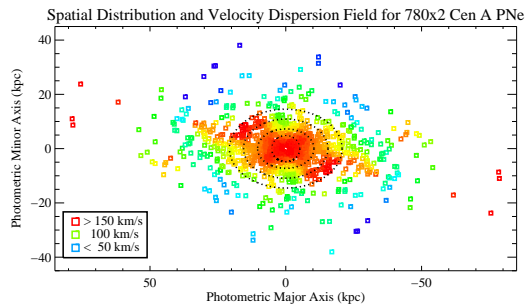}
\caption[\cena\ velocity dispersion field]
{\small \cena\ velocity dispersion field.  The PN sample
used was antisymmetrized.  The dispersion was calculated using a
variable bandwidth kernel with the bandwidth set to the larger of 3~kpc or
the distance to the tenth nearest neighbor.  Dispersion estimates beyond
$\sim5r_e$ are likely to be biased and should be ignored.  The colors
are determined by the magnitude of the dispersion, with red points
representing the dynamically hottest regions.  In the inner
regions, there is a large degree of pressure support which declines
gradually along the major axis, and more quickly along the minor axis.
\label{figure:rvs_vdisp}
}
\end{figure}

\subsection{Intrinsic Shape and Triaxiality}

Observationally, two-dimensional velocity data in the halos of
early-type galaxies are still rare.  When such data sets exist, they are
useful for comparing with models and simulations.  In particular, we
focus on the predicted velocity fields for triaxial galaxies and merger
remnants.

Elliptical galaxies have long been suspected to be triaxial in shape
(Binney 1978; Illingworth 1977).  However, recovering their true axial
ratios is difficult from imaging alone because we can only see their
shapes projected onto the sky.  Many have shown that mapping the
two-dimensional velocity field gives additional leverage on this problem
(e.g.\ Binney 1985; Wilkinson \etal 1986; Statler 1987).
In an attempt to constrain the intrinsic shapes
of elliptical galaxies, Statler (1991, S91) studied the morphology of
projected velocity fields as a function of axial ratios and viewing
angles.  In a triaxial potential, stars can follow ``tube'' orbits about
the long and short axes, or ``box'' orbits that have no net angular
momentum (there are others, but we limit this
discussion to these families).  Intermediate ($y$\,) axis tube orbits, however,
are not stable.  As a result, the distribution of $x$-tube and $z$-tube
orbits, in conjunction with the viewing angle, determine the shape of
the velocity field.

For certain viewing angles, both kinematic
misalignment and kinks in the zero-velocity contour will result.
Examples of this are shown in S91's Figure~3.
Some of these
velocity fields are qualitatively similar to the \cena\ field in 
Figure~\ref{figure:rvs_zvc}.  While this particular model is not
necessarily the best for comparing with \cena---S91 used more
extreme axial ratios to emphasize the effect---it is a nice illustration
of how the ZVC can change with viewing angle in a triaxial galaxy.

H95 combined PNe kinematics with the orientation of the dust lane to
constrain the orientation and axial ratios of \cena.  By assuming
that the inner portion of the 
central gas disk has settled and lies in the plane perpendicular to
the long axis, they were able to determine that we are viewing \cena\
nearly along the intermediate axis.  Conveniently, this means that the
photometric major axis corresponds to the intrinsic long axis, and the 
photometric minor axis corresponds to the intrinsic short axis.  The
viewing angle that they derived was $\theta=90\degr$ and $\phi=107\degr$.

Like in S91, $\theta$ is defined as the angle of the line of sight
from the $z$-axis.  Since we are supposedly looking down
a line-of-sight (LOS) that is in the $x-y$ plane, this value is $90\degr$.
A possible point of confusion is that 
Hui and Statler define $\phi$ differently.  Hui defines it as the
angle in the $x-y$ plane of the LOS from the long $x$-axis, while Statler
defines it as the angle from the $-y$-axis.  The result is that 
$\phi_{H95} = 107\degr$ is equivalent to 
$\phi_{Statler} = 107\degr+90\degr = 197\degr$.
This same LOS can equivalently 
be described as $\theta=-90\degr$ and $\phi_{Statler}=17\degr$.

With fewer PNe, H95 measured a single kinematic axis in \cena\ that was
misaligned by $39\degr$ from the photometric minor axis.  
As a result, Mathieu, Dejonghe, \& Hui (1996) classified the
morphology of the velocity field as ``Nn'' --- a normal, single
maximum/minimum velocity field with a normal core whose outer ZVC
deviates from the core ZVC by less than $30\degr$.  
With our larger catalog and denser
sampling, we see that the ZVC does in fact deviate significantly in the
core.  Hence, we reclassify \cena's velocity field as ``Dd'', i.e.\ one with
two peaks in velocity (seen in this case along the major and minor axes)
and a kinematically distinct core.  
This is consistent with S91's
statement that D-type fields and d-type cores are most favored when the
view is down the $y$-axis. 

The non-linearity of the ZVC 
also affects H95's determination of \cena's axial ratios.  
If $a$, $b$, and $c$ correspond to the long, intermediate,
and short axes, and the axial ratios are defined as $\zeta = b/a, 
\xi=c/a$ and $1 \geq\zeta\geq\xi$, then
the triaxiality of a galaxy can be parameterized as
\begin{equation}
T = \frac{1-\zeta^2}{1-\xi^2}
\end{equation}
(S91).  $T$ is equal to 0 for an oblate spheroid and 1 for a
prolate spheroid. 
The axial ratios can then be expressed as
\begin{equation}
\zeta^2 = 1 - \frac{T(1-e^2)}{1-Te^2cos^2\phi}
\end{equation}
\begin{equation}
\xi^2 = 1 - \frac{1-e^2}{1-Te^2cos^2\phi}
\end{equation}
where $e$ is the observed (projected) axial ratio, and which we take to
be 0.8 (H95).  

The asymptotic position angle of the ZVC with respect to the intrinsic
short axis, which in the case of \cena\ is the photometric minor axis,
depends on the degree of
triaxiality as $sin^{-1}\sqrt{T}$ (S91).  
Because H95 had a smaller sample of
PNe in a smaller area, they determined that this angle was $39\degr$.  
However, as we have seen from
Section~\ref{section:pnvfield}, this angle is
actually closer to $83\degr$.  Using this new value, we obtain
$T=0.985$, a value that is
significantly more prolate than the $T=0.40$ derived by H95.  Given
the previous values for the viewing angles, we obtain axial ratios 
$(b/a,c/a) = (0.791,0.787)$.  While the value of $c/a$ is almost
identical to H95's, our value for $b/a$ is smaller than H95's value of
0.92.

Whereas H95 found \cena\ to be slightly oblate and significantly triaxial, 
we find that the shape of \cena\ is very nearly prolate.  
Statler's velocity field classifications (his Figure~6),
however, show that a galaxy this prolate is unlikely to show a ``Dd''
velocity field from this viewing angle.  This discrepancy could arise
because of uncertainty in the angle of the ZVC.
Nevertheless, this clearly shows the importance of measuring the
velocity field well into the halo.  Only the dense sampling and large
spatial extent of our survey showed the more precise asymptotic angle
of the ZVC.  
Of course, this treatment assumes a certain degree of symmetry and
dynamical relaxation.  While we find that the velocity field is quite
symmetric, the outer isophotes of \cena\ still show the effects of
non-equilibrium motions that must have been caused by recent
interactions.  
Merger simulations might help shed more light on the
intrinsic shape of the potential, and whether figure rotation might be
necessary to reproduce the observed kinematics.

\subsection{Mass Estimates}
\label{section:mass}

One of the major advantages of using discrete velocity tracers such as
PNe is that they are detectable at very large radii.  Because we can
only constrain the mass within our most distant test particle, getting
kinematic information in the halos of galaxies is essential for
determining their total mass.  Total masses of elliptical galaxies have
been estimated dynamically using spectral absorption-line profiles 
(e.g.\ Gerhard \etal 2001, Graham \etal 1999), 
PNe (e.g.\ H95, Ciardullo \etal 1993),
globular clusters (e.g.\ C\^{o}t\'{e} \etal 2001, Romanowsky \& Kochanek
2001, C\^{o}t\'{e} \etal 2003), hot X-ray gas (Forman, Jones, \& Tucker 1985), 
weak gravitational lensing (McKay \etal 2002), as well as
combinations of these techniques.  For the most part, these observations
support the existence of substantial dark matter halos around
ellipticals, with the exact nature of the outer regions a function of
environment.  Since gravity dominates most formation processes, we
expect that many properties of galaxies
should scale with their total gravitational mass.  Therefore, it is
important to measure the total masses of nearby galaxies, where the data
is obtainable.  Using the smaller sample of PNe, H95 estimated the mass of
\cena\ within 25~kpc to be $3.1\times10^{11} M_{\sun}$ with a
mass-to-$L_B$ ratio of 10.
We use the PNe in \cena\ to estimate its
total dynamical mass within 80~kpc and compare this to previous work.

Dynamical mass estimation in pressure-supported systems such as
elliptical galaxies is more complicated than for disk galaxies.  With only
two dimensions of spatial 
information, and one dimension of velocity data, assumptions about the
symmetry of the galaxy and the isotropy of the velocity ellipsoid are
necessary.  One of the more convenient family of mass estimators uses the
weighted mean of $v_{los_i}R_i/G$ where $v_{los_i}$ is the line of sight
velocity and $R_i$ is the projected radius.  The exact weighting factor
depends on the assumptions the various estimators make.  Variants on this theme
include the virial mass estimator, the projected mass estimator
(Heisler, Tremaine, \& Bahcall 1985), and the tracer mass estimator (TME)
by Evans \etal (2003).  The latter is designed for the use of a ``tracer''
population, such as PNe or globular clusters, 
whose spatial distribution does not match
the overall mass distribution.  Simulations show that this estimator
does at least as well as the other estimators, and has significantly
less bias.  Given these well-suited properties, we chose to use the
tracer mass estimator.

For this mass estimator, the mass enclosed within the outermost tracer
particle is
\[ M = \frac{C}{GN} \sum_i{v_{los_i}^2. R_i} \]
where
\[ C = \frac{4(\alpha+\gamma)}{\pi} \frac{4-\alpha-\gamma}{3-\gamma} 
\frac{1-(r_{in}/r_{out})^(3-\gamma)}{1-(r_{in}/r_{out})^(4-\alpha-\gamma)}, \]
the form of gravitational force field is $\psi(r) \propto r^{-\alpha}$,
and the space density of tracer particles goes as $r^{-\gamma}$.  This
mass estimate is only for pressure support, so the contribution from rotation
must be added separately.  

We estimate $\gamma$ using the surface density of PNe in circular
annular bins, shown in Figure~\ref{figure:tme_radialpn}.  
The inner 5~kpc shows the effects of
incompleteness, and the drop off at large radii is likely due to the
increased diskiness of the isophotes.  However, the PNe between
projected radii of 5 and 40~kpc nicely follow a power law with $\gamma =
2.54$.  In addition, we assume an isothermal potential ($\alpha=0$) and an
isotropic velocity distribution.  These are both shown to be reasonable
assumptions for halo velocity tracers in nearby galaxies, e.g.\ M~31
(Evans \etal 2003), M~87 (C\^{o}t\'{e} \etal 2001), and M~49
(C\^{o}t\'{e} \etal 2003).

\begin{figure}
\plotone{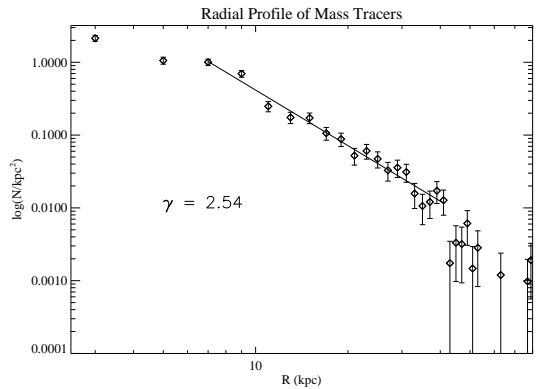}
\caption[Radial Distribution of PNe]
{\small Radial distribution of PNe.  We fit a power-law to the radial surface
density of PNe ($\Sigma\propto R^{\gamma}$) for the purpose of using the
tracer mass estimator.  PNe 
were binned in circular annuli.  The inner regions show the effects of
incompleteness, while the drop off in the outer regions is likely due to
breakdown of our assumption of spherical symmetry.
\label{figure:tme_radialpn}}
\end{figure}

The TME only calculates the contribution of random motions to the mass.
As we have shown, the stars in \cena\ have a large rotational component
that must be included.  This contribution is separable in the Jeans
equation and is simply $M_{rot}(r) = \langle v_{rot}\rangle^2 r/G$, where
$\langle v_{rot}\rangle$ is the mean rotation speed.  
We take $\langle v_{rot}\rangle$ to be
100~\kms\ since this is where the PN rotation curve flattens.  The
rotation of the PNe will be discussed more extensively in the following
section.

The total dynamical mass of \cena\ within 80~kpc, as estimated by the
tracer mass estimator, is $5.3{\pm0.5}\times10^{11} M_{\sun}$.  The
error in this estimate is purely statistical and was derived using the
bootstrap with 1000 resamplings.  Assuming that
the stellar component of \cena\ has a $B$-band luminosity of
$L_B = 3.98\times10^{10} L_{\sun}$ (H95), we get $M/L_B = 13\pm1$.  

For self-consistency, we compare masses from the TME with previously derived
values.  When we use only the PNe available to H95, we derive a mass
within 25~kpc of $3.3\pm0.6\times10^{11} M_{\sun}$, a value in close agreement
with the mass of $3.1\times10^{11} M_{\sun}$ derived in H95 using the
spherical Jeans equation, and that of
$\sim4\times10^{11} M_{\sun}$ estimated in Mathieu, Dejonghe, \& Hui
(1996) using the QP technique.

\subsection{The Extended Stellar Disk}

The presence of an extended stellar disk along the major axis allows us
to make some useful simplifications.  As did H95, we treat this major
axis sample as a thick, rotating disk in a spherical potential.
Although we have evidence that \cena's potential is likely
triaxial in shape, the approximation of spherical symmetry in this case
is not likely to introduce egregiously large biases because \cena\ is
not very triaxial.  In fact, when Mathieu \etal (1996)
modeled \cena\ using triaxial St{\"a}ckel potentials, their mass estimate was 
statistically indistinguishable from the value obtained from the
spherical Jeans equation.

\begin{figure*}
\plotone{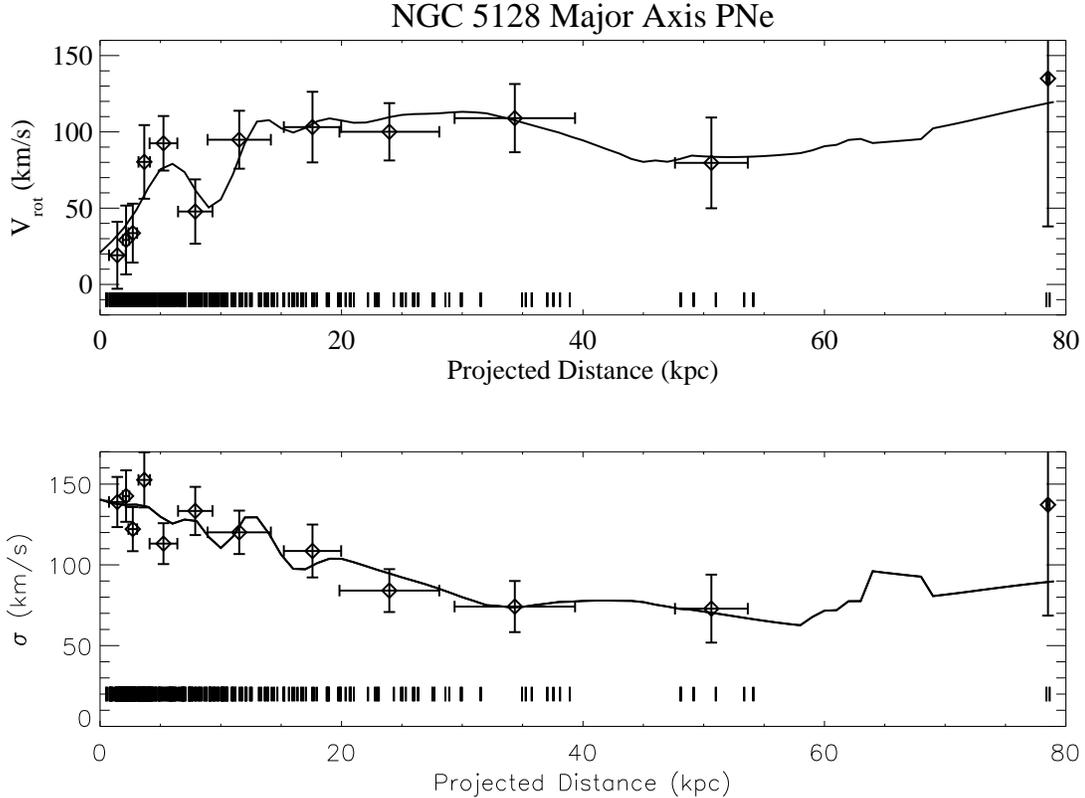}
\caption[PN Major Axis Rotation Curve and Velocity Dispersion Profile]
{\small Rotation curve and velocity dispersion for major axis PNe.  We use the
PNe along the extended stellar disk to show the rotation and dispersion
in the stars.  PNe included are either within a perpendicular distance
of $\pm2\arcmin$ from the major axis or $\pm10\degr$ cone centered on the
major axis.  In both plots, the points represent 
binned values along the major axis while the solid line is the local
kernel-smoothed value where the kernel width is a function of the local
density.  
(a) The top panel shows that the
rotation curve rapidly rises and flattens at $\sim100$~\kms.  (b) The
bottom panel shows the dispersion dropping from a central value of
140~\kms to a minimum of $\sim75$~\kms.  Beyond 20~kpc, $V/\sigma$ varies
between 1 and 1.5.
\label{figure:oned_pn}}
\end{figure*}

In Figure~\ref{figure:oned_pn}, we show the rotation curve and
line-of-sight velocity
dispersion profile for PNe along the major axis.  PNe were included in
this sample if they had a perpendicular distance from the
major axis of $\pm2\arcmin$ {\it or} were within a $\pm 10^{\degr}$ cone
about the major axis with respect to the origin.  The first criterion
is the same as for H95, and the latter was added so that more PNe from
the outer regions could be included.  The PNe on one side
were reflected through the origin to create a single radial profile for
302~PNe.  The points with error bars in Figure~\ref{figure:oned_pn} are
the binned values as a function of radius.  These values are listed in
Table~\ref{table:oned_pntable}.  We also kernel smoothed the
entire data set with a variable-width gaussian, with the width inversely
proportional to the local density of PNe.  The errors in $V_{rot}$ and
$\sigma$ are equal to $\sigma/\sqrt{N}$ and $\sigma/\sqrt{2N}$,
respectively.

\begin{deluxetable}{rrr}
\tablecaption{Binned Major Axis PNe Rotation and Dispersion}
\tablewidth{0pt}
\tablehead{
\colhead{$R_{proj}$} & \colhead{$V_{rot}$} & 
\colhead{$\sigma$} \\
\colhead{(kpc)} & \colhead{\kms} & \colhead{\kms}
}
\startdata
 1.4 & $  19\pm 22$ & $ 139\pm 16$ \\
 2.2 & $  29\pm 23$ & $ 143\pm 16$ \\
 2.7 & $  34\pm 19$ & $ 122\pm 14$ \\
 3.7 & $  80\pm 24$ & $ 153\pm 17$ \\
 5.3 & $  93\pm 18$ & $ 113\pm 13$ \\
 7.9 & $  48\pm 21$ & $ 133\pm 15$ \\
11.5 & $  95\pm 19$ & $ 120\pm 13$ \\
17.6 & $ 103\pm 23$ & $ 109\pm 16$ \\
24.0 & $ 100\pm 19$ & $  84\pm 13$ \\
34.4 & $ 109\pm 22$ & $  74\pm 16$ \\
50.6 & $  80\pm 30$ & $  73\pm 21$ \\
78.5 & $ 135\pm 97$ & $ 137\pm 69$ \\
\enddata
\label{table:oned_pntable}
\end{deluxetable}

Figure~\ref{figure:oned_pn}a shows that the rotation velocity of major axis
PNe rises until 10~kpc, at which point it stays flat at $\sim100$~\kms.  
This behavior was seen out to 25~kpc in H95, and appears to be constant
out to the limits of our data.  The velocity dispersion has a central
value of $\sim140$~\kms, consistent with absorption line spectroscopy
(Wilkinson \etal 1986), and slowly declines to $\sim75$~\kms\ at 50~kpc.
Outside of 20~kpc, the ratio between rotational and pressure support,
$V/\sigma$ is between 1 and 1.5.

\subsection{Mass Estimation with the Jeans Equation}
Following the methods of H95, we apply the spherical Jeans equation to
the major axis rotation and 
line-of-sight velocity dispersion profile to derive the dynamical mass
distribution in \cena.  Briefly, the Jeans equation for a spherical,
isotropic stellar system is 
\begin{equation}
\frac{d(\rho\sigma^2)}{dr} = - \frac{GM(r)\rho}{r^2} + 
\frac{\rho V^2_{rot}}{r}
\end{equation}
which, when solved for the velocity dispersion is
\begin{equation}
\sigma^2(r) = \rho^{-1} \int_{r}^{\infty}\frac{GM(x)\rho}{x^2}dx -
	\rho^{-1}\int_{r}^{\infty} \frac{\rho V_{rot}^2}{x}dx
\end{equation}

\begin{figure*}
\plotone{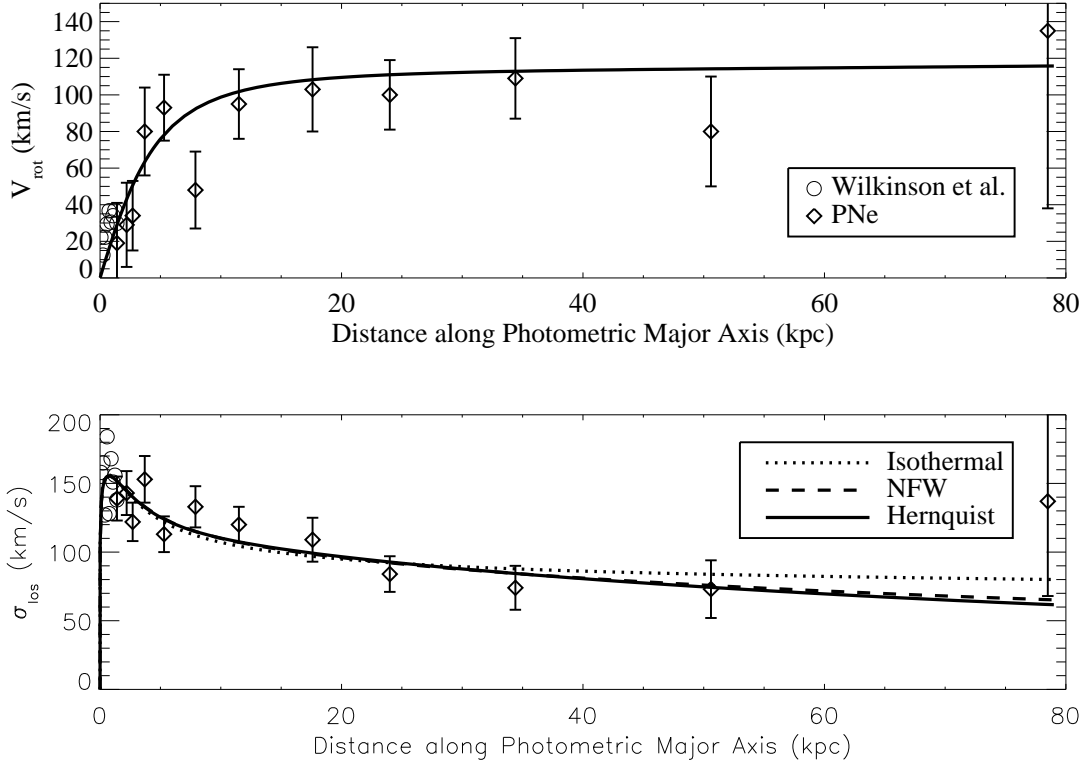}
\caption[Jeans Equation fit to the Major Axis PNe]
{(a) Major axis rotation curve.  We fit a parametrized flat rotation curve to
the rotation data shown in 
Figure~\ref{figure:oned_pn}a.  (b) Jeans Equation fit to the major axis
velocity dispersion profile.  In the bottom panel is the best fit
line-of-sight velocity dispersion profile for three different
dark halo mass models.
\label{figure:jeansplot}}
\end{figure*}

Another assumption we must make is the form of the mass model.  
We adopt a two-component mass model for the luminous and dark matter in
the galaxy.  Like
H95, we choose the Hernquist (1990) mass model for the luminous
component, which has the form
\begin{equation}
\rho(r) = \frac{M_l a}{2\pi} \frac{1}{r(r+a)^2},
\end{equation}
where $M_l$ is the total luminous mass, and $a$ is the scale length and
relates to the effective radius of the familiar de Vaucouleurs law as 
$a = r_e/1.8153$.

For the dark component, we assume three different mass
models: the Hernquist mass profile, a pseudo-isothermal halo with a core, 
and the mass profile from Navarro,
Frenk, \& White (1996, 1997, NFW).  The density profile of the isothermal halo
has the form
\begin{equation}
\rho(r) = \frac{\rho_0}{1+(r/D)^2},
\end{equation}
where $\rho_0$ is the central density, and $D$ is the scale length.
Correspondingly, the cumulative mass profile is
\begin{equation}
M(r) = 4\pi\rho_0 D^2\ [r - D\ {\rm arctan}(r/D)].
\end{equation}
The NFW density and mass profiles take the forms
\begin{equation}
\rho(x) = \frac{\rho_s}{x (x+1)^2},
\end{equation}
\begin{equation}
M(x) = 4\pi\rho_s r_s^3\ [{\rm ln}(1+x) - \frac{x}{1+x}],
\end{equation}
where $x=r/r_s$, $r_s$ is the scale radius at which the slope of the
profile transitions to the $r^{-3}$ regime, and $\rho_s$ is the
density at $r_s$.
Thus, the total mass model is 
\begin{equation}
M(r) = \frac{M_l r^2}{(r+a)^2} + M_{dark}(r),
\end{equation}
where $M_l$ is the luminous mass of the galaxy which
is allowed to vary, $a$ is the scale length which is fixed from the
galaxy surface photometry, and $M_{dark}(r)$ is one of the three mass
models described above.

We also parametrize the intrinsic (three-dimensional) rotation curve as
\begin{equation}
V_{rot}(r) = \frac{v_0 r}{\sqrt{(r^2 + r_0^2)}},
\end{equation}
which asymptotically approaches the value of $v_0$ as
$r\rightarrow\infty$.  We numerically integrate these functions over the
line-of-sight to obtain the projected rotation, surface mass
profile, and line-of-sight velocity dispersion profile.  

In the fitting procedure, we first fit for $v_0$ and $r_0$ from the rotation
curve data.  Then, we solve for the best fit luminous and dark mass
parameters given the 
projected velocity dispersion profile data.  For the luminous component,
we assume and fix $r_e = 5.185$~kpc (Dufour \etal 1979).  The three
free parameters in the mass model
are then $M_l$, a dark matter scale length ($d$, $D$,
$r_s$), and a mass or density scale ($M_d$, $\rho_0$, $\rho_s$).

\begin{deluxetable*}{lcccccc}
\tablewidth{0pt}
\tabletypesize{\scriptsize}
\tablecaption{Mass Model Fits \label{table:massmodels}}
\tablehead{
\colhead{Model} & \colhead{$M_l$} & 
\colhead{Scale Mass/Density} &
\colhead{Scale Radius} &
\colhead{$M_{tot}(r < 80~{\rm kpc})$} &
\colhead{$M_l/L_B$} &
\colhead{$M_{tot}/L_B$} \\
\colhead{(1)} &
\colhead{(2)} &
\colhead{(3)} &
\colhead{(4)} &
\colhead{(5)} &
\colhead{(6)} &
\colhead{(7)}
}
\startdata
Isothermal & $1.5\times10^{11}$ &   0.0243 &   4.5 & $5.9\times10^{11}$ &   3.8 &  15 \\
NFW & $1.4\times10^{11}$ &   0.010 &  14.4 & $5.1\times10^{11}$ &   3.4 &  13 \\
Hernquist & $1.4\times10^{11}$ & $7.2\times10^{11}$ &  31.7 & $5.0\times10^{11}$ &   3.5 &  12 \\
\tablenotetext{1}{Dark Matter halo model}
\tablenotetext{2}{Total luminous mass ($M_{\sun}$)}
\tablenotetext{3}{$\rho_0$ for isothermal ($M_{\sun}/pc^3$), $\rho_s$
for NFW ($M_{\sun}/pc^3$), $M_d$ for Hernquist ($M_{\sun}$)}
\tablenotetext{4}{$D$, $d$, and $r_s$, in kpc}
\tablenotetext{5}{Total mass within 80 kpc}
\tablenotetext{6}{Mass-to-light of luminous matter}
\tablenotetext{7}{Total mass-to-light within 80 kpc}
\enddata
\end{deluxetable*}

Figure~\ref{figure:jeansplot} shows the best-fit rotation curve and
dispersion profiles.  The best-fit parameters for the rotation curve are
$v_0 = 141$~\kms  and $r_0 = 6.3$~kpc.  
In these plots and fits, we also include binned long-slit
spectroscopic kinematic data from Wilkinson \etal (1986) at distances
within $1\farcm5$.  These data fill a region in which there are few PNe
due to the high background and large extinction toward the center of the
galaxy.  

We show the results from our model fits in
Table~\ref{table:massmodels} and in the velocity dispersion plot of
Figure~\ref{figure:jeansplot}.  All three mass models give equally acceptable
fits to the data.  At small radii, the isothermal halo has a core while
the other models have $r^{-1}$ cusps.  At large radii,
the isothermal halo falls off as $r^{-2}$, the NFW as $r^{-3}$,
and the Hernquist profile as $r^{-4}$.  This behavior is reflected
in the fitted
parameters, where the isothermal halo requires a larger luminous mass,
smaller scale length, and gives the largest total mass.  The NFW and
Hernquist profiles are similar, except that the Hernquist model requires
a larger scale length because of its steeper falloff at large radii.
For the NFW halo, we also calculate the concentration parameter, 
$c = r_{200}/r_s$, where $r_{200}$ is the ``virial radius'', defined to
be the radius inside of which the average density is 200 times the
critical density.  The halo concentration for our best-fit NFW halo
model is $c=12$, which is consistent with values obtained from simulations
(NFW 1997).

While comparisons between these three density profiles is an interesting
exercise, we are not able to use our data to differentiate between them.
It may in principle be possible to rule out one or more with much higher
quality data at large radii, but the lack of kinematic tracers, and 
various degeneracies with orbital anisotropy and the
assumed light distribution make the task difficult.

All three models give total
masses within 80~kpc of 5--6$\times10^{11} M_{\sun}$, with 
$M_{tot}/L_B\sim$12--15.  The parameters determined for the Hernquist
dark halo are not significantly different from those determined in
H95 using only PNe within 25~kpc.  Moreover, these values are in good
agreement with those obtained with the tracer mass estimator.

An interesting check on these mass estimates is to see how many
PNe, if any, would be unbound from the galaxy given the estimated mass
profile.  In Figure~\ref{figure:escapev}, we plot the radial velocity of
each PN versus its projected radius.  The overplotted solid lines
mark the escape velocity at that true radius given the best-fit
two-component Hernquist mass model.  The escape velocity takes the form
\begin{equation}
v_{esc} = \sqrt{2G(\frac{M_l}{r+a} + \frac{M_d}{r+d})}.
\end{equation}
The dotted lines mark the escape
velocity if there is no dark matter.  Many PNe are
not bound to the galaxy if there is no dark matter
(however, this is only obvious if the PN sample extends beyond 10~kpc).
This shows once again that a dark component is necessary to explain the
stellar velocity distribution.  

All the PNe, even the outermost ones,
are bound by our best-fit mass model.  Although it is true that since we
make the assumption of equilibrium, most of the PNe should appear bound,
this does not preclude any one PN from having $V > v_{esc}$.  It is of
interest that the most distant stars in \cena\ have velocities that are
still consistent with being bound to the galaxy.  This would not
necessarily be the case for galaxies in a cluster environment.  
Intracluster PNe have been identified in the Virgo and Fornax clusters,
and have velocities that trace the cluster potential (Arnaboldi \etal
2002).  An increase in $\sigma$ could also be caused by a
recent tidal interaction with a neighbor.
The fact that \cena\ is in a loose group of galaxies with a lack of
any nearby neighbors may help explain why these most distant PNe have
very regular velocities. 

\begin{figure}[t]
\plotone{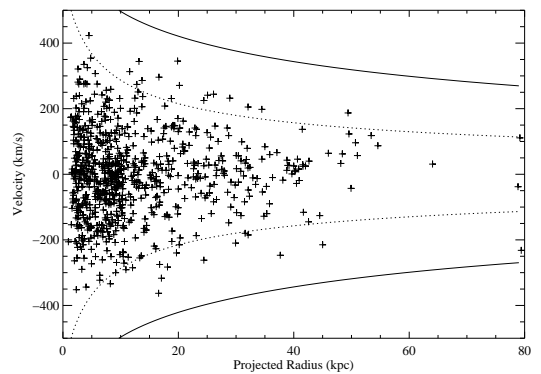}
\caption[Escape Velocity versus Radius]
{\small Escape velocity versus radius for PNe.  Each point marks the projected
radius and radial velocity of a PN.  The solid lines are the local
escape velocity for our best-fit two-component Hernquist mass model.
The dotted lines are the same, except for a model with no dark matter.
Without dark matter, many PNe are unbound.  With dark matter, even the
most distant PNe are consistent with being bound.
\label{figure:escapev}}
\end{figure}

\section{Discussion}
\subsection{Comparison to Merger Simulations}

N-body simulations of disk-disk mergers suggest that spheroidal stellar
populations may be the result of one or more mergers. Moreover,
Barnes (1992) showed that many simulated merger remnants have misaligned
kinematic axes. Recent
simulations are producing two-dimensional velocity fields that are
useful comparisons for our data.  In particular, we refer to the work of
Bendo \& Barnes (2000) and Naab \& Burkert (2001).

Both of these groups simulated the mergers of two stellar disk/bulge
systems with
different mass ratios and angular momentum vectors at varying
angles.  It is important to note that these studies did not include gas
dynamics or star formation, so cannot reproduce a remnant such as \cena\
in detail.  However, the general trends in stellar kinematics are useful
baselines for comparison.
They tested two families of mergers---``equal-mass'' (1:1) and
``unequal-mass'' (3:1).  All merger remnants have masses of
3--5$\times10^{11} M_{\sun}$, values which are comparable to the total
mass of \cena. 
Equal-mass mergers tend to produce a wide variety of
kinematic properties, often disrupting many of the initial
characteristics of the progenitor galaxies.  Unequal-mass mergers tend
to preserve the disk of the larger galaxy, and hence will more
consistently have rotational support.

All kinds of mergers can produce the types of velocity field morphology
we see in \cena.  These simulations show that the projected
velocity contours can kink, much like those in Statler's triaxial
galaxies, and in our observed velocity field.  The remnant's final
kinematics will be the result of the initial angular momentum
(both rotational and orbital) of the progenitor galaxies, and the shape
of the final potential.

The properties of \cena\ are similar to some aspects of both types of
mergers.  It is important to keep in mind that even if we were certain
that \cena\ was the product of a 
disk-disk merger, it would be extremely difficult to uniquely constrain
the exact parameters of the merger that occurred.
Given the diskiness of the outer isophotes in \cena, and the rapid
rotation along the major axis, it is possible that \cena\ is the result
of an unequal-mass disk-disk merger.  On the other hand, 
the severe misalignment of
the kinematic axis in the halo is more typical of 1:1 mergers.  Naab \&
Burkert (2001) use simulations to show that even these 1:1 remnants 
may contain a disk component, although these simulated disks are much
smaller in size than the one observed in \cena.  An
interesting note is that the 3:1 simulated merger remnants in 
Bendo \& Barnes (2000) with
kinematic properties most like the 1:1 merger remnants were also the most
prolate.  These prolate remnants also showed a relatively large
population of X-tube orbits, and these may be responsible for the
twist in the ZVC.  Thus, in the simplest comparison that we can make
with these simulations, the morphology and kinematics of \cena\ are
broadly consistent with the remnant of an unequal-mass merger.

Two-dimensional velocity fields have the potential to tell us much more
about the structure and formation of ellipticals if the corresponding
models and simulations are available.  Simulations that also include gas
dissipation and star formation would be especially useful.
Ongoing surveys such as SAURON and the PN spectrograph promise to
accumulate a wealth of data in the coming years.

\subsection{Mass-to-Light}

By assuming isotropy, we derive a mass within 80~kpc of 
$5\times10^{11} M_{\sun}$, which gives a mass-to-light ratio of
$M/L_B\sim13$.  We can also compare this mass to the total K-band
luminosity, which should be a better measure of the old stellar
population of the galaxy.  Recently, the Two Micron All Sky Survey
(2MASS) imaged and measured
photometry for \cena\ in their Large Galaxy Atlas (Jarrett \etal 2003).
They measured an apparent total magnitude of $K_s = 3.942\pm0.016$, which
translates to $M_{K_s} = -23.778$, a value slightly brighter than
$M^{*}_K$ for the $K$--band galaxy luminosity function (Kochanek \etal
2001).  This gives a $K_s$-band luminosity of  $L_{K_s} =
6.97\times10^{10} L_{\sun}$ and a mass-to-light ratio  of $M/L_{K_s}\sim7$.

The reader must be warned that mass estimates are fraught with
systematic uncertainties due to the many assumptions made.  For
instance, we have assumed that the surface radial distribution of stars
continues as either a power law, de Vaucouleurs law, or Hernquist
profile even at very large radii.  Given the extremely low surface
brightness of the integrated stellar light, and the small number of PNe
at these distances (we detect 4 PNe beyond 60~kpc), it is currently
extremely difficult to accurately constrain the density profile at these
radii.  However, the amount of light in these regions is negligible for
the purposes of determining the total luminosity of the galaxy.
Likewise, the abundance of two-dimensional velocity data
behooves us to eventually model the mass distribution with 
non-spherical geometries.  In this case, we are fortunate that the
unique, nearly edge-on disk-like component in \cena\ works in our favor by
simplifying the geometry of the tracer PNe in the halo.  What we can see
is that the stars in \cena\ do exist in detectable numbers at distances
as great as 80~kpc (at least in one direction), and that their
velocities are fairly well-behaved, extending the flat rotation profile
seen in the inner halo.

Although dark matter is required to explain the observed stellar
kinematics in \cena, our value for
$M/L_B$ is noticeably low when compared to those determined for
ellipticals using X-ray halos and satellite kinematics.  This is
interesting, especially in light of recent results from the PN
spectrograph group that report low mass-to-light ratios for ellipticals
within 5~$r_e$ (Romanowsky \etal 2003).  Bahcall, Lubin, \& Dorman (1995) 
compiled the values of $M/L_B$ for galaxies in the literature and found
that they are closer to $200\pm50\ R/(100{\rm kpc})$ for ellipticals and 
$60\pm10\ (R /100{\rm kpc})$ for spirals.  If we assume $H_0 = 70$ and
calculate $M/L_B$ for $R=80$~kpc, then
these are lowered somewhat to values of $112\pm28$ and $34\pm6$, 
but both are still larger than what we are measuring by over
$3\sigma$. H95 also remarked that the
dynamical mass they measured within 25~kpc was systematically lower than
that measure by Forman, Jones, \& Tucker (1985) from ROSAT data.
van Gorkom \etal (1990)
reached a similar conclusion when comparing \ion{H}{1} dynamical mass
estimates with the X-ray estimates.
However, Kraft \etal (2003) recently used Chandra and XMM data 
to measure the total gravitating mass with 15~kpc and determined a value
of $\sim2\times10^{11} M_{\sun}$, which is statistically
indistinguishable from the values we obtain from our mass models within
the same radius.  This independent mass determination implies
that our assumptions of orbital isotropy
and a high inclination for the stellar disk are not likely to be grossly
wrong, at least within 15~kpc.  

Could it be possible that \cena\ appears brighter than expected, 
either because of
an error in measurement or because it is young?  Using our value of the
gravitating mass, \cena\ would have to be over two magnitudes
fainter than what Dufour \etal (1979) measured in order for $M/L_B$ to
be at the level of 80--100.  One potential culprit is that 
the obscuring dust lane makes
extrapolation toward the central regions essential for obtaining a
magnitude that one can compare with other galaxies.  Hence, there will
always be a larger than normal uncertainty in the galaxy's measured
luminosity, although this is less true for infrared wavelengths.  
However, we think it is unlikely that such a large error was made.
Could \cena\ be intrinsically brighter than most
ellipticals because it is young?  Using the 2001 release of the Bruzual
\& Charlot (1993) evolutionary synthesis models, we find that for solar
metallicity, the mean age of the ``old'' component of \cena\ (excluding
the dust lane region) would need to have an age of 2~Gyr in order to be
two magnitudes brighter than a similarly enriched 12~Gyr old population.
While there is growing evidence that some fraction of the stars in
\cena\ formed in a secondary event (e.g.\ Rejkuba 2003), it is unlikely 
the entire galaxy
could be this young, especially since some globular clusters in the
galaxy that are at least 10~Gyr old (discussed in Peng, Ford, \& Freeman
2004c).  Additionally, color-magnitude diagram studies of the red giant
population in the halo show no evidence for a dominant young component
(Harris \& Harris 2002).

\subsection{Comparison with Other Ellipticals}

While some of the most massive ellipticals can have central velocity
dispersions of $\sigma\sim300$~\kms, \cena\ has a more intermediate
value for its central velocity dispersion with $\sigma\sim150$~\kms.
Likewise, with absolute magnitudes of $M_B = -20.8$ and $M_{K_s} = -23.778$,
\cena\ has an typical luminosity for early-type galaxies.   
The measured amount of total luminous mass is
$1.4\times10^{11} M_{\sun}$, and its stellar mass-to-light ratio is
$M_l/L_B\sim3.5$.  
These properties of \cena's stellar component all are what one would
expect for a typical elliptical galaxy, so there is little reason to 
suspect that its dark halo should be less massive than in other
galaxies.  While it is possible that our
assumption of isotropy for the stellar orbits is incorrect, 
fully radial orbits (which maximizes the discrepancy)
might increase the total $M/L$ by a factor of two, but not a factor of ten.

Could the low value for $M/L$ be explained by the luminous component
rather than the total mass (i.e.\ by the stellar population rather than
the dark matter)?  If so, then this should be evident in the colors of
the galaxy. 
How do the mean integrated optical colors of \cena\ compare to those of
other, perhaps less morphologically disturbed ellipticals?
We compare the \uv\ color of \cena\ 
to those of Virgo and Coma elliptical galaxies using photometry from
Bower, Lucey, \& Ellis (1992).  The colors of ellipticals are known to
correlate with galaxy properties such as velocity dispersion, surface
brightness, and luminosity so it is important to compare physically
similar galaxies.  We plot the color-velocity dispersion relation for
these galaxies and for \cena, because the velocity dispersion for \cena\
is well-measured with a value of 150~\kms\ (Wilkinson~\etal 1986, H95).  
We have obtained our own measure of the colors of \cena\ using
CCD photometry of the galaxy introduced in Peng \etal (2002), 
and further described in Peng, Ford, \& Freeman (2004b).
We deredden our measured colors assuming
$E(B-V) = 0.115$ (Schlegel, Finkbeiner, \& Davis 1998) 
to obtain $(U-V)_0 = 1.28$.  These results are
plotted in Figure~\ref{figure:colvdisp}.  We also plot the significantly
bluer measurement for \cena\ of van den Bergh (1976, vdB76).

\begin{figure}
\epsscale{1.0}
\plotone{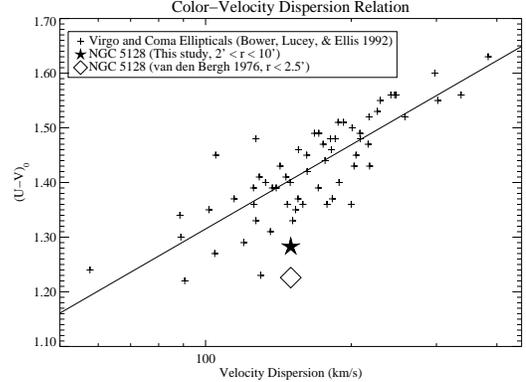}
\caption[(\uv) Color-Velocity Dispersion Relation for Elliptical
Galaxies with \cena]
{\small (\uv) Color-Velocity Dispersion Relation for Elliptical
Galaxies with \cena.  We compare our measurement of the mean (\uv) color
and velocity dispersion of \cena\ with those measured for Virgo and Coma
early-type galaxies from Bower, Lucey, \& Ellis (1992).  The value from
van den Bergh (1976) is bluer than ours, probably because that color is
for a more central region.  \cena\ appears significantly bluer than the
cluster ellipticals for its velocity dispersion, although this may due
to metallicity gradients.\label{figure:colvdisp}}
\end{figure}

The difference between our measurement and the value from vdB76
underline the importance of aperture effects.  vdB76 measured the color
of the galaxy out to a radius of $2\farcm5$, a region that is still
affected by the young stars in the dust lane.  vdB76 notes that the
color of the galaxy becomes bluer as one measures regions closer to the
nucleus.  We chose to generously mask the dust lane to avoid any light
from obviously recent star formation.  Our measurement is for a region
2--10 arc minutes along the major axis, and 5--10 arc minutes along the
minor axis.  It is not surprising that our measured color is redder than
that measured by vdB76.

In Figure~\ref{figure:colvdisp}, we also see that \cena\ is fairly blue
for its velocity dispersion compared to Virgo and Coma ellipticals.
Even our redder measurement falls below the color-velocity dispersion
relation.  One could argue that this may also be an aperture effect.
Elliptical galaxies are known to have metallicity (and hence, color)
gradients.  The aperture through which Bower \etal measured their
photometry has a radius of $2.5 h^{-1}$~kpc, which is 1.75~kpc for
$H_0 = 70$.  Measurements through an aperture of this size for \cena\
would be dominated by the dust lane and its associated star formation.
If we assume, however, that \cena\ does have an obscured, redder stellar
population comparable to those in the cluster ellipticals, we can
make some assumptions on the nature of the color gradient and
estimate what its color would be.  Franx \& Illingworth (1990) 
measured the mean $(U$--$R)$ color gradient in ellipticals to 
be $-0.23$~mag per
decade of radius.  Since most of this is due to the change in (\ub), we
assume this value as an upper limit on the gradient in (\uv).  The mean
radius at which we measure our color is $\sim6$~kpc.  If we assume that
the mean radius of the Bower \etal measurements are $\sim1$~kpc, then the
maximum expected color difference between our values would be 0.18~mag.
Adding this value to our (\uv)$_0$ color, for a value of 1.46, does in
fact bring \cena\ to the mean of the color-velocity dispersion
relation.  The major problem with this extrapolation is that we do not see
color gradients in \cena\ of this magnitude, although our measurements
are still preliminary.  So, while we measure the color of the ``old''
stellar light in \cena\ to be significantly bluer than the central
regions of local ellipticals at the same velocity dispersion, 
we will need to do further work on color gradients to determine whether
this is because these galaxies are intrinsically different, or because
the apertures of the measurements are different.

If the colors are intrinsically different, this 
could be another clue that there exists an
intermediate-age component in the galaxy.  A recent star-formation event
of only 10\% by mass can significantly change the optical colors of
otherwise old stellar populations.  However, a difference of only
0.12~mag between \cena\ and the mean Virgo-Coma relation cannot account
for the difference in age of 10~Gyr needed to make up the $M/L_B$
``deficit''.

We can compare this mass-to-light value with the work of 
Romanowsky \etal (2003), who recently reported that the three
ellipticals in their study are not dominated by dark matter within
5~$r_e$.  For NGC~3379, they find that the
ratio of dark matter to total mass within 5~$r_e$ (9~kpc) is less than
0.32.  The situation in \cena\ is not very different.  For \cena, 
this ratio within 9~kpc is 0.26, within 5~$r_e$ is
0.55, and within 80~kpc is 0.73.  So, we find that the transition from
luminous- to dark matter-dominated enclosed mass occurs at approximately
5~$r_e$.  

It is possible that much of the dark matter is beyond 80~kpc, and thus
not detected by our kinematic tracers.  \cena\ looks to be the product of
an unusual merger --- systems with such extended disks are somewhat 
unique.  It may be that the merger parameters which led to such an
extended disk also led to the dark halo being even more extended.  If
that is true, the mass measured by the PNe and GC, and also by the X-ray
data, would not include much of the dark matter.  However, what is
perhaps a bit puzzling is that the dark matter models that we fit to the
current data do not require a low concentration, which might be expected
if the dark halo was very diffuse.

\section{Summary}

We extended the original PN survey of H95 by surveying three times
farther into the halo.  There are now 1141 known PNe in \cena, of which
780 are confirmed with radial velocities.  Many of these new PNe are at
projected distances beyond 20~kpc.

Despite the disturbed appearance of \cena\ in optical images, the
velocity field of the PNe is very regular.  The point symmetry of the
potential is evidenced in the velocity field.  There is a high degree of
rotation along the major axis that coincides with a faint stellar disk.
Beyond 20~kpc, this disk has a flat rotation curve at a velocity of
100~\kms.  There also exists significant minor axis rotation, resulting
in a well-defined twist in the line of zero velocity.  This ``S''-shaped
zero velocity contour is characteristic of triaxial potentials and
merger remnants.

Using this larger sample of PNe, we re-derive the intrinsic axial ratios
of \cena\ and find that the galaxy may be nearly prolate, with axial ratios
$b/a \approx c/a \approx 0.79$.  This is in contrast to the more oblate
shape derived by H95.  However, this estimate is at odds with other
model predictions which require a more triaxial shape, and merits
further study.  We also find that the kinematics are consistent 
with simulated disk-disk merger remnants, and could perhaps be the
result of one with a 3:1 mass ratio.

The entire PN sample is used to estimate the dynamical mass of the
galaxy using the tracer mass estimator.  We also use
the kinematics of the major axis stellar disk to derive a mass from the
spherical Jeans equation.  The rotation curve is flat out to the limit
of our data while the velocity dispersion profile falls gradually.
Taken together, the TME and our best-fit Hernquist,
isothermal halo, and NFW 
models give an average mass within
80~kpc of $5\times10^{11} M_{\sun}$, and $M/L_B\sim13$.  
These models produce masses
that agree with determinations from the latest X-ray data.
As H95 showed, dark matter is
necessary to fit the velocity distribution of PNe beyond $\sim10$~kpc.

The mass-to-light ratio of \cena\ is found to be abnormally low when
compared to those determined for other ellipticals.  It is unlikely that
this $M/L_B$ difference is either due to an incorrect measurement of the
galaxy light, or because the galaxy is extremely young.  It may be that
the dark matter halo is very extended, and that a large fraction is
beyond 80~kpc.
We defer further discussion of the formation of \cena\ to a future
paper (Peng, Ford, \& Freeman 2004c), in which we will compare these field
star kinematics with those of the globular clusters.

\acknowledgments

E.\ W.\ P.\ acknowledges support from NSF grant AST 00-98566. 
H.\ C.\ F.\ acknowledges support from NASA contract NAS 5-32865
and NASA grant NAG 5-7697. We thank the staffs at CTIO and the AAO for their
invaluable help during our observing runs.  We also thank David Malin for
making available to us his deep photographic prints of NGC 5128. 
E.\ W.\ P.\ acknowledges Tom Statler for useful discussions on the
zero-velocity contour, Roger Peng for his insight on non-parametric
smoothers, and Eric Barnes for discussions on mass models.  We thank the
anonymous referee for constructive comments.
This research has made use of the NASA/IPAC Extragalactic
Database (NED), which is operated by the Jet Propulsion Laboratory,
California Institute of Technology, under contract with NASA.

\end{document}